\documentclass[superscriptaddress,aps,preprintnumbers,amsmath,amssymb,prd,nofootinbib,preprint]{revtex4-1}
\pdfoutput=1

\usepackage{graphicx}
\usepackage{epstopdf}
\usepackage{dcolumn}
\usepackage{bm}
\usepackage{bbm}
\usepackage{hyperref}
\usepackage[usenames, dvipsnames]{color}
\usepackage{amsmath,amssymb}
\usepackage[sort&compress]{natbib}
\usepackage{float}
\usepackage{multirow}
\usepackage{slashed}
\usepackage{cancel}
\usepackage[table]{xcolor}
\usepackage{multirow}
\usepackage{booktabs}
\usepackage{makecell}
\usepackage{mathtools}
\usepackage{pifont}
\usepackage{enumitem}
\usepackage{setspace}
\usepackage{ulem}

\newcommand{\eV}{\textrm{ eV}}

\newcommand{\GeV}{\textrm{ GeV}}

\newcommand{\brr}[1]{\left(#1\right)}
\newcommand{\srr}[1]{\left[#1\right]}

\numberwithin{equation}{section}

\makeatletter
\renewcommand{\p@subsection}{}
\renewcommand{\p@subsubsection}{}
\makeatother

\hypersetup{colorlinks,citecolor= blue,linkcolor= blue}

\begin{document}

\setstretch{1.1}

\preprint{UMN-TH-4326/24, CETUP2024-003}

\title{A Light QCD Axion with Hilltop Misalignment}

\author{\small Raymond T. Co}
\affiliation{\footnotesize Physics Department, Indiana University, Bloomington, Indiana 47405, USA}
\affiliation{\footnotesize School of Physics and Astronomy, University of Minnesota, Minneapolis, Minnesota 55455, USA}
\affiliation{\footnotesize William I. Fine Theoretical Physics Institute, University of Minnesota, Minneapolis, Minnesota 55455, USA\vspace{1.5cm}}

\author{\small Tony Gherghetta}
\affiliation{\footnotesize School of Physics and Astronomy, University of Minnesota, Minneapolis, Minnesota 55455, USA}

\author{\small Zhen Liu}
\affiliation{\footnotesize School of Physics and Astronomy, University of Minnesota, Minneapolis, Minnesota 55455, USA}

\author{\small Kun-Feng Lyu}
\affiliation{\footnotesize School of Physics and Astronomy, University of Minnesota, Minneapolis, Minnesota 55455, USA}

\begin{abstract}
We study the cosmological evolution of a light QCD axion and identify the parameter space to obtain the correct relic dark matter abundance. The axion potential is flattened at the origin, corresponding to the only minimum, while it is unsuppressed at $\pi$. These potential features arise by assuming a mirror sector with the strong CP phase $\bar\theta$ shifted by $\pi$ compared to the SM sector, which allows the mirror axion potential to be tuned against the usual QCD axion potential. Before the QCD phase transition, assuming the mirror sector is decoupled and much colder than the SM thermal bath, the mirror sector potential dominates, causing the axion to initially roll to a temporary minimum at $\pi$. However, after the QCD phase transition, the potential minimum changes, and the axion relaxes from the newly created ``hilltop'' near $\pi$ to the CP-conserving minimum at the origin. As the axion adiabatically tracks this shift in the potential minimum through the QCD phase transition, with non-adiabatic evolution near $\pi$ and 0, it alters the usual prediction of the dark matter abundance. Consequently, this ``hilltop" misalignment mechanism opens new regions of axion parameter space, with the correct relic abundance while still solving the strong CP problem, that could be explored in future experiments.

\end{abstract}

\maketitle

\tableofcontents


\section{Introduction}

The QCD axion is a simple and elegant solution to the strong CP problem~\cite{Peccei:1977hh, Peccei:1977ur,Weinberg:1977ma,Wilczek:1977pj}. The strong CP-violating phase $\bar\theta$ is promoted to a dynamical field $\phi_a$, the axion, identified as the pseudo Nambu-Goldstone boson that arises from the spontaneous breaking of Peccei-Quinn symmetry at the decay constant scale, $f_a$. The axion field couples to the QCD topological term with a suppression scale proportional to $f_a$. 
A potential for the axion is induced via this coupling by nonperturbative dynamics after QCD confines.
This explicit breaking of the Peccei-Quinn symmetry causes the axion field to relax to a field value that minimizes the total potential, where $\theta_a \equiv \phi_a/f_a$ cancels the strong CP phase, $\bar{\theta}$.
The QCD axion mass can be precisely calculated via chiral perturbation theory and is given by
\begin{equation}\label{eq:qcd_axion_mass}
   m_{a,\text{QCD}}^2 = \dfrac{m_u m_d}{(m_u + m_d)^2} \dfrac{m_\pi^2 f_\pi^2}{f_a^2}  \  ,
\end{equation}
where $m_\pi$ and $f_\pi$ are the pion mass and decay constant, respectively and $m_{u,d}$ are the up, down quark masses.
A more precise result is computed by Ref.~\cite{Gorghetto:2018ocs} by considering higher-order corrections.
There is an active worldwide experimental effort searching for the QCD axion which has yet to probe a large fraction of the parameter space in the $(1/f_a, m_a)$ plane~\cite{ParticleDataGroup:2022pth}.
 
In most of the $(1/f_a, m_a)$ parameter space  the axion does not address the strong CP problem. Therefore, one interesting question is whether the axion mass could deviate from the conventional QCD prediction Eq.~\eqref{eq:qcd_axion_mass} while still solving the strong CP problem. Indeed, there are a number of different scenarios which can lead to an axion that is either heavier or lighter than the value in Eq.~(\ref{eq:qcd_axion_mass}). To enhance the axion mass, QCD must be modified at UV scales and various approaches have been proposed including exploiting small instanton effects from a UV broken gauge group~\cite{Dimopoulos:1979pp,Tye:1981zy,Holdom:1982ex,Holdom:1985vx,Gherghetta:2016fhp,Agrawal:2017ksf,Gaillard:2018xgk,Csaki:2019vte, Gherghetta:2020ofz,Valenti:2022tsc}, an extra dimension~\cite{Gherghetta:2020keg}, or from imposing a $\mathbb{Z}_2$ symmetry~\cite{Rubakov:1997vp,Berezhiani:2000gh,Hook:2014cda,Dimopoulos:2016lvn,Hook:2019qoh,Dunsky:2023ucb}.

On the other hand, suppressing the QCD axion mass appears much more challenging. Currently, there are only two viable models. One is the proposal by Hook~\cite{Hook:2018jle} (and further studied in Ref.~\cite{DiLuzio:2021pxd, DiLuzio:2021gos}), where the axion mass can be naturally suppressed by imposing a discrete $\mathbb{Z}_N$ symmetry. By summing up the potential contributions from the QCD sector plus $N-1$ copies of the mirror QCD sector, assuming all the potentials have the same magnitude but shifted $\theta$ angles, the total axion potential turns out to be suppressed by $(m_u/m_d)^{N}$. The second possibility is the so-called anarchic QCD axion~\cite{Elahi:2023vhu}. Based on the DFSZ axion model, the phase of the $B_\mu$ term can shift the physical theta angle. By tuning the size of the potential coefficients, the physical axion field can obtain a smaller mass. 

Apart from the particle physics model building, the cosmological evolution of the axion field has also received much attention. 
The axion is well-known to serve as the dark matter candidate~\cite{Preskill:1982cy,Abbott:1982af,Dine:1982ah}. One of the most significant approaches to produce the correct dark matter abundance is the misalignment mechanism. 
The axion field is initially misaligned from the minimum of the potential. 
Due to the cosmic expansion, the axion can only start to evolve when the Hubble expansion rate drops down to the axion mass. 
This misalignment mechanism only happens in 
a specific region of the axion parameter space, if the initial misalignment angle is of order unity.
There have been proposals of various production mechanisms to extend the allowed parameter space, such as a modified cosmological background~\cite{Hashimoto:1998ua,Visinelli:2009kt,Kawasaki:2011ym,Baer:2011eca,Bae:2014rfa,Co:2016xti}, parametric resonance~\cite{Co:2017mop,Harigaya:2019qnl},  kinetic misalignment~\cite{Co:2019jts}, and dynamical misalignment~\cite{Dvali:1995ce,Banks:1996ea,Daido:2017wwb,Co:2018phi,Co:2018mho,Takahashi:2018tdu,Takahashi:2019pqf,DiLuzio:2021gos,Heurtier:2021rko}. 

In this paper, we will focus on models with a mirror sector where the $\bar\theta$ angle is shifted by $\pi$  
compared to the Standard Model (SM) sector. This causes the mirror axion potential to have an opposite sign compared to the usual QCD axion potential and therefore the axion mass can be reduced by tuning the amplitude of the mirror axion potential. We will consider two examples for the new physics potential. The first example relies on the nonperturbative dynamics of mirror QCD to generate a potential identical in form to the usual QCD axion potential, except that the mirror Higgs vacuum expectation value (VEV) is tuned to reduce the axion mass. Alternatively, in the second example, the mirror QCD group is assumed to be spontaneously broken by a newly introduced mirror QCD charged scalar. A cosine potential is then generated by instantons where the VEVs of the mirror Higgs doublet and QCD charged scalar are tuned to cancel the QCD contribution to the axion potential. In both examples, the height of the total potential at $\theta_a=\pi$ is not significantly suppressed and the tuning is connected to scalar field VEVs in the mirror sector. Since the Higgs mass hierarchy is not yet understood, we remain agnostic about how the scalar field VEVs are obtained.

Given that there are examples, albeit tuned, where the QCD axion potential can be flattened near the origin but not at $\pi$, we next investigate the cosmological consequences of an axion potential with these features. The cosmological evolution is similar to the dynamical misalignment~\cite{Co:2018mho} and the trapped misalignment~\cite{DiLuzio:2021gos} mechanisms. In the early universe, before the QCD phase transition, the axion potential is assumed to be dominated by the mirror QCD contribution which has a minimum at $\theta_a = \pi$.
Thus, the axion field first oscillates and relaxes to $\pi$. Only when the SM QCD contribution to the axion potential becomes significant does the potential minimum switch from $\theta_a = \pi$ to $\theta_a = 0$. This is in contrast to the $\mathbb{Z}_N$ model~\cite{Hook:2018jle,DiLuzio:2021gos}, where there are $N$ distinct degenerate minima of the potential and the axion can only reside in the desired CP conserving minimum $\theta_a = 0$ with a $1/N$ probability. In our case, we have a unique minimum at $\theta_a = 0$, so that the axion always evolves to the correct vacuum required to solve the strong CP problem.  
Moreover, we numerically track the time-dependent potential in greater detail compared to Ref.~\cite{DiLuzio:2021gos} and discover distinct features where the axion adiabatically rolls down to the origin,  with non-adiabatic evolution near $\pi$ and the origin.
This new behavior in the axion field evolution points to different parameter space to obtain the correct dark matter abundance, thereby providing further motivation to search for lighter QCD axions.

This paper is organized as follows. In Section~\ref{sec:formalism}, we briefly introduce  examples of models with a light QCD axion and discuss the features of the axion potential as well as effects at finite temperature and constraints from cosmology. A detailed analysis of the cosmological evolution is then given in Section~\ref{sec:cosevol} where the equation of motion is solved and the axion relic abundance is calculated. Finally, we conclude in Section~\ref{sec:conclusion}.


\section{Reducing the Axion Mass}\label{sec:formalism}

We aim to reduce the axion mass without spoiling the solution to the strong CP problem. This requires a cancellation in the axion potential which can be achieved by requiring that 
the second derivative of the potential contribution from the new physics sector is negative at $\theta_a = 0$.
Hence, the origin must be a local maximum for the potential contribution generated by the new physics.  
One way to obtain a negative potential contribution is to introduce a mirror QCD sector where $\bar{\theta}$ is effectively shifted by $\pi$.
This can be realized by imposing a $\mathbb{Z}_2$ exchange symmetry 
which is non-linearly realized on the axion, $\theta_a\rightarrow \theta_a +\pi$~\cite{Hook:2018jle,DiLuzio:2021pxd}. Explicitly in the KSVZ scenario, with a complex scalar field $\Phi= \frac{\varphi}{\sqrt{2}} e^{i\theta_a}$ (where $\langle \varphi \rangle = f_a$) and KSVZ fermions $\Psi_{L,R}$ plus the mirror fermions $\Psi'_{L,R}$, we can have the following terms
\begin{equation}
    \mathcal{L} \supset - y \Phi \bar{\Psi}_L \Psi_R + y'\Phi \bar{\Psi}'_L \Psi_R' + \text{h.c.}\,,
\end{equation}
which is invariant under the following $\mathbb{Z}_2$ symmetry\footnote{Note that this $\mathbb{Z}_2$ symmetry differs from the $\mathbb{Z}_2$ symmetry considered in Ref.~\cite{Hook:2019qoh}, where $\Phi\rightarrow \Phi$, which was then used to {\it increase} the axion mass.}
\begin{equation}
\text{SM} \leftrightarrow \text{SM}'\quad; \quad\quad
    \Phi \leftrightarrow
    - \Phi\,,
\end{equation}
provided the Yukawa couplings satisfy $y=y'$ and where prime ($^\prime$) denotes fields/couplings in the mirror sector. After integrating out the heavy KSVZ fermions,
the axion $\phi_a$ couples to the two sectors via
\begin{equation}
    \mathcal{L} \supset \brr{\dfrac{\phi_a}{f_a} + \bar{\theta}}\dfrac{\alpha_s}{8\pi} G^{\mu\nu} \widetilde{G}_{\mu\nu} + \brr{\dfrac{\phi_a}{f_a} + \bar{\theta} + \pi} \dfrac{\alpha_s}{8\pi} G'^{\mu\nu} \widetilde{G}'_{\mu\nu}\,,
\end{equation}
where $\alpha_s=g^2/4\pi$ is the QCD fine structure constant, $\widetilde{G}_{\mu\nu}=\frac{1}{2} \epsilon_{\mu\nu\rho\sigma}G^{\rho\sigma}$ with $G^{\mu\nu}$ the gluon field strength and similarly for the mirror sector.
Nonperturbative dynamics in both the QCD and mirror sectors then induces the total axion potential
\begin{equation}
    V_{\rm total}(\phi_a,\bar{\theta}) = V_{\rm QCD}(\phi_a,\bar{\theta}) + V_{\rm new}(\phi_a,\bar{\theta} + \pi)\,,
    \label{eq:Vtot}
\end{equation}
where 
\begin{equation}\label{eq:QCD_axion_potential}
    V_{\rm QCD}(\phi_a,\bar{\theta}) =  - m_{\pi}^2 f_\pi^2 \sqrt{1-\frac{4 z}{(1+z)^2} \sin^2\left( \frac{\phi_a}{2 f_a} + \frac{\bar{\theta}}{2} \right)}=- m_{\pi}^2 f_\pi^2 +\frac{1}{2} m_{a,\rm QCD}^2 \phi_a^2 +\dots\,, 
\end{equation}
with $z \equiv m_u/m_d$. 
The second equality in Eq.~\eqref{eq:QCD_axion_potential} is obtained by expanding the shifted axion field around the global minimum at the origin $\phi_a\rightarrow \phi_a +\langle \phi_a\rangle =\phi_a  -\bar \theta f_a$, which gives rise to the axion field mass term with $m_{a,\rm QCD}$ given in Eq.~\eqref{eq:qcd_axion_mass}.

The contribution $V_{\rm new}$ in Eq.~\eqref{eq:Vtot} is the potential generated by the new physics sector. In order to solve the strong CP problem and have a lighter QCD axion mass the potential must have the following features.
First, the potential $V_{\rm new}$ is assumed to have the same period as $V_{\rm QCD}$ so that there is only one unique minimum for $V_{\rm total}$ located at $\bar{\theta} = 0$. 
This is in contrast to the $\mathbb{Z}_N$ model in Refs.~\cite{Hook:2018jle,DiLuzio:2021pxd}, where there are $N$ degenerate minima located at $\bar{\theta} = 2\pi k/N$ with $k \in \mathbb{Z}$. This avoids a $1/N$ selection accident of $N$ degenerate minima for the $\mathbb{Z}_N$ models to solve the strong CP problem. The second feature is that the amplitude of the potential $V_{\rm new}$ must be comparable to the amplitude of the QCD potential in order for there to be a cancellation. Thus, expanding Eq.~\eqref{eq:Vtot} about the origin gives
\begin{equation}
    V_{\rm total}\supset \frac{1}{2}(m_{a,\rm QCD}^2 -m_{a,\rm QCD'}^2)\phi_a^2 +\dots\equiv \frac{1}{2} \varepsilon m_{a,\rm QCD}^2\phi_a^2 +\dots\,,
    \label{eq:totalpot}
\end{equation}
where $m_{a,\rm QCD'}^2$ is the contribution to the axion mass from the mirror QCD sector.
The QCD axion mass is then $m_a^2\equiv \varepsilon m_{a,\rm QCD}^2$, where 
\begin{equation}
    \varepsilon \equiv 1-\dfrac{m_{a, \rm QCD'}^2}{m_{a,\text{QCD}}^2} \, .
    \label{eq:epsdefn}
\end{equation}
Clearly, to obtain a light QCD axion, the mirror contribution $m_{a, \rm QCD'}$ must be tuned to be comparable to the QCD contribution $m_{a, \rm QCD}$. Furthermore, we emphasize that the tuning parameter must be chosen so that $\varepsilon>0$, in order to have a global minimum at the origin. If $\varepsilon <0$,   the global minimum shifts to $\theta_a = \pi$ (with the origin becoming a maximum) and the strong CP problem is no longer solved.
In the next subsection, we will present two examples to achieve this cancellation.

\subsection{Examples from a mirror sector}
\label{sec:examples}

We will consider two possible types for the potential $V_{\rm new}$ in Eq.~\eqref{eq:Vtot}. Both of these scenarios will require an explicit breaking of the $\mathbb{Z}_2$ symmetry and a tuning in the potential. These tunings will be related to the size of scalar field VEVs in the mirror sector.

\subsubsection{Explicitly broken $\mathbb{Z}_2$ model via mirror Higgs}
\label{sec:Z2mirrorHiggs}

If the mirror sector were an exact copy of the SM where, in particular, the mirror Higgs VEV $v'$ is equal to the SM Higgs VEV $v$, then as discussed in~\cite{DiLuzio:2021pxd}, the origin of the total axion potential is a local maximum. Instead, we will consider the possibility that the $\mathbb{Z}_2$ symmetry is explicitly broken by assuming $v'\neq v$, while keeping all other dimensionless parameters equal. Note that this explicit $\mathbb{Z}_2$ breaking via the Higgs VEVs is soft and only logarithmically affects marginal operators where, in particular, the effects on $\bar \theta$ are negligible~\cite{Hook:2019qoh}.
The mirror Higgs VEV $v'$ can then be tuned to obtain a local minimum and a cancellation between the two axion mass contributions.
After the mirror QCD confines, the axion potential in the mirror sector is given by
\begin{equation}
    V_{\rm new}^{(\rm conf.)}(\phi_a,\bar{\theta} + \pi) =  - m_{\pi'}^2 f_{\pi'}^2 \sqrt{1-\frac{4 z}{(1+z)^2} \sin^2\left( \frac{\phi_a}{2 f_a} + \frac{\bar{\theta} + \pi}{2} \right)} \,,
    \label{eq:potential_new_1}
\end{equation}
where $m_{\pi'}$ and $f_{\pi'}$ is the mirror pion mass and decay constant, respectively. 
Using the Gell-Mann-Oakes-Renner relation, $m_{\pi'}^2 f_{\pi'}^2$ is proportional to $(m_{u'}+m_{d'}) \Lambda_{\text{QCD}'}^3$, where $m_{u',d'}=y_{u,d} v'$ are the mirror quark masses with identical Yukawa couplings $y_{u,d}$ as in the SM sector. Taking the second derivative of Eq.~\eqref{eq:potential_new_1}, evaluated at $\phi_a+\bar{\theta}f_a=0$, gives a tachyonic mass, assuming $z<1$, with the magnitude
\begin{equation}
    m_{a,\text{QCD}'}^2 = \dfrac{m_{\pi'}^2 f_{\pi'}^2}{f_a^2} \dfrac{z}{1-z^2}  \,. \\
    \label{eq:maQCDp}
\end{equation}
Note that $m_{a,\text{QCD}'}^2 = \frac{1+z}{1-z}  \frac{d^2}{d\phi_a^2} V_{\rm new}^{(\rm conf.)}\big|_{\phi_a+\bar{\theta}f_a=\pi}$,
which implies that the second derivative at the origin of the shifted potential compared to that of the unshifted potential is larger by a factor of $(1+z)/(1-z) \approx 3$ (giving rise to a local maximum at $\phi_a+\bar{\theta}f_a=0$, when $v=v'$). Hence, requiring that Eq.~\eqref{eq:maQCDp} is equal to $-(1-\varepsilon) m_{a,\text{QCD}}^2$ (using Eq.~\eqref{eq:epsdefn}), gives the condition
\begin{equation}
    m_{\pi'}^2 f_{\pi'}^2 = \brr{1 - \varepsilon} \dfrac{1-z}{1+z} m_{\pi}^2 f_{\pi}^2 \ .
    \label{eq:piontuning}
\end{equation}
Using Eq.~\eqref{eq:piontuning}, the light QCD axion mass becomes $m_a^2=\varepsilon m_{a,\text{QCD}}^2$ where
\begin{equation}
    \varepsilon= 1 - \dfrac{(1+z) v' \Lambda_{\text{QCD}'}^3}{(1-z)v {\Lambda_{\text{QCD}}^3 }} \approx 1 - 3 \brr{\dfrac{v'}{v}}^{5/3}\,.
    \label{eq:eps}
\end{equation}
To obtain the second expression in Eq.~\eqref{eq:eps}, we have used $z \approx 1/2$ and solved the one-loop renormalization group equation to obtain $\Lambda_{\text{QCD}'}/\Lambda_{\text{QCD}}\simeq(v'/v)^{2/9}$ for $v'<v$. Thus, to obtain $\varepsilon <1$, Eq.~\eqref{eq:eps} implies that $v'/v$ should be tuned to be slightly smaller than $3^{-3/5} \simeq 0.52$. In this way, the mirror Higgs doublet VEV $v'$ provides the tuning parameter to obtain a light QCD axion mass. Clearly, a mirror SM sector is constrained in the early Universe, and these effects will be discussed in Section~\ref{sec:thermal}.

\subsubsection{Spontaneously broken mirror QCD}
\label{sec:spontbrokenmirrorQCD}

An alternative possibility for $V_{\rm new}$ is to assume a cosine potential that can arise from a dilute instanton gas calculation. For example, one can introduce a massive QCD-charged real adjoint scalar $S$ (with mass $m_S \gtrsim 10$ TeV) in the SM sector and a QCD$'$ charged scalar $S'$ in the mirror sector where both fields have the same potential (i.e. $m_S=m_S'$ and identical quartic couplings), except that the mass-squared is positive (negative) for $S (S')$.
Consequently, only $S'$ obtains a non-vanishing VEV $v'_S$, which spontaneously breaks the mirror QCD group. This means that small mirror instantons of size $\rho\lesssim 1/v'_S$ will generate a cosine potential for the axion. 

To obtain a sizeable mirror instanton contribution to the axion potential, we will assume that the mirror QCD breaking scale $v'_S$ lies below the mirror $u$ quark mass $m_{u'}$, which requires $v'\gg v'_S$.
Since all mirror quarks are much heavier than $v'_S$, there is no chiral suppression in the instanton amplitudes~\cite{Novikov:1984ecy}. The instanton potential is then given by~\cite{tHooft:1976snw,Csaki:2019vte}
\begin{equation}\label{eq:V_instanton}
V^{(\rm inst.)}_{\rm new}(\bar{\theta},\theta_a+\pi) = - e^{-i \brr{\bar{\theta} + \theta_a+\pi}} \!\!\!\int \dfrac{d \rho}{\rho^5} \dfrac{C_1 e^{-\alpha(\frac{1}{2})}}{(N-2)! (N-1)!} \!\!\brr{\dfrac{8\pi^2}{g^{\prime 2}(1/\rho)}}^{\!\!\! 2 N} \!\!\!\! e^{-\frac{8\pi^2}{g^{\prime 2}(1/\rho)} - C_2 N - 2\pi^2 \rho^2 v_S^{\prime 2}} + \text{h.c.}\,,
\end{equation}
where $N=3$, $C_1 \approx 0.466,C_2 \approx 1.678$ and $\alpha(\frac{1}{2})=0.145873$. 
Note that the term $-2\pi^2 \rho^2 v_S^{\prime 2}$ in the exponential factor of Eq.~\eqref{eq:V_instanton} is the effect of the spontaneously broken mirror QCD symmetry to the instanton action and serves as a natural cutoff for the integration over the instanton size $\rho$. 
The scale $v_S^\prime$ depends on the specific symmetry breaking mechanism. The $SU(3)$ mirror QCD group can be broken to $U(1) \times U(1)$ with an adjoint scalar $S'$ or can be completely broken if more scalars are introduced. We do not specify the details and simply use $v_S^\prime$ as a parameterization of the symmetry breaking mechanism.
Due to the $\mathbb{Z}_2$ symmetry, the QCD gauge coupling $g$ is the same as the mirror QCD gauge coupling $g'$ at UV scales. However, given the assumed large hierarchy $v' \gg v$, the $\mathbb{Z}_2$ symmetry is broken at lower energy scales and therefore the couplings only remain equal until $v'$ i.e. $g(v') = g'(v')$. Below the scale $v'$, the running of the two gauge couplings will deviate. In the mirror sector, the heavier mirror quarks lead to a larger $\beta$-function value compared to the 
QCD $\beta$-function at the same scale. 
Hence, the mirror gauge coupling at the scale  $v'_S$ can be much larger than $g(v'_S)$, thereby reducing the amount of exponential suppression in Eq.~\eqref{eq:V_instanton}. The one-loop running of the mirror gauge coupling is
\begin{equation}
    \dfrac{1}{\alpha'(\mu)} = \dfrac{1}{\alpha'(v'_S)} + \dfrac{b'}{2\pi} \log\dfrac{\mu}{v'_S} \,,
    \label{eq:gcoupling}
\end{equation}
where $\mu$ is the renormalization scale and $b'$ is the $\beta$-function coefficient that depends on the matter content. Substituting Eq.~\eqref{eq:gcoupling} into Eq.~\eqref{eq:V_instanton} we can analytically integrate over $\rho$ to obtain the approximate result
\begin{equation}
    V^{(\rm inst.)}_{\rm new}(\bar{\theta},\theta_a+\pi) \approx 2^{1-\frac{b'}{2}} \pi^{4-b'} \Gamma\brr{\dfrac{b'}{2} - 2} C_1 e^{-\alpha(\frac{1}{2})- 3 C_2} \brr{\dfrac{8\pi^2}{g^{\prime 2}\brr{v'_S}}}^{6} e^{-\frac{8\pi^2}{g^{\prime 2}(v'_S)}}
     v_S^{\prime 4}\, \cos\brr{\theta_a+\bar{\theta}+\pi}\,,
    \label{eq:approxint}
\end{equation}
where $\Gamma$ denotes the gamma function. Note that the amplitude of the cosine potential is proportional to $v_S^{\prime 4}$ since the mirror QCD group is spontaneously broken and does not confine (i.e. $\Lambda_{\rm QCD'} < v'_S$ assuming that $v_S=v'_S\gtrsim 10$ TeV and $v'\gtrsim 10^9$ GeV). 
The gauge coupling $g'(v'_S)$ depends on the mirror Higgs VEV $v'$ and using the fact that $b' = 21/2$ (assuming one adjoint real scalar field) near the scale $v'_S$, Eq.\eqref{eq:approxint} can be numerically approximated as
\begin{equation}
    V^{(\rm inst.)}_{\rm new}(\bar{\theta},\theta_a+\pi) \approx 
    -(86\, \text{MeV})^{4} 
    \brr{\dfrac{v'}{10^{11}\,\text{GeV}}}^{4} \brr{\dfrac{v'_S}{30\, \text{TeV}}}^{-7} \cos\brr{\theta_a+\bar{\theta} + \pi}\,,
    \label{eq:analVinst}
\end{equation}
where for simplicity we have assumed the charged scalar mass $m_{S'}=v'_S$. 
An exact computation gives a result that differs from the approximation in Eq.~\eqref{eq:analVinst} by a factor of 1.6. 
Requiring that the amplitude of the cosine potential in Eq.~\eqref{eq:analVinst} matches $(1-\varepsilon) m_{a,\text{QCD}}^2 f_a^2$, in order to cancel\footnote{Note that if the $\mathbb{Z}_2$ symmetry acts on the axion as $\theta_a\rightarrow \theta_a$ then the instanton-induced cosine potential has the same sign as the QCD potential and can be used to {\it increase} the axion mass, similar to Ref.~\cite{Hook:2019qoh}.} the QCD contribution to the potential, we obtain
\begin{equation}
    \varepsilon \simeq 1 -  \brr{\dfrac{v'}{10^{12}\,\text{GeV}}}^{4} \brr{\dfrac{v'_S}{130\, \text{TeV}}}^{-7}\,.
    \label{eq:epseqn2}
\end{equation}
Thus, we see that to obtain a value $\varepsilon < 1$ we can tune the values of $v'$ and $v'_S$. 
We require that the charged scalar mass is above 10 TeV, but  smaller than the mirror $u$-quark mass. Interestingly, for the scalar mass range $10~{\rm TeV}\lesssim m_{S'} \lesssim 3\times 10^4~{\rm TeV}$ a reduced axion mass can be obtained for the mirror Higgs VEV $10^{10}~{\rm GeV}\lesssim v' \lesssim 10^{16}~{\rm GeV}$. 
Again, there will be cosmological constraints on the mirror sector and these will be discussed in Section~\ref{sec:thermal}.

\subsection{Locally-flat axion potential} 

\begin{figure}[t!]
    \centering
    \includegraphics[width=0.8\linewidth]{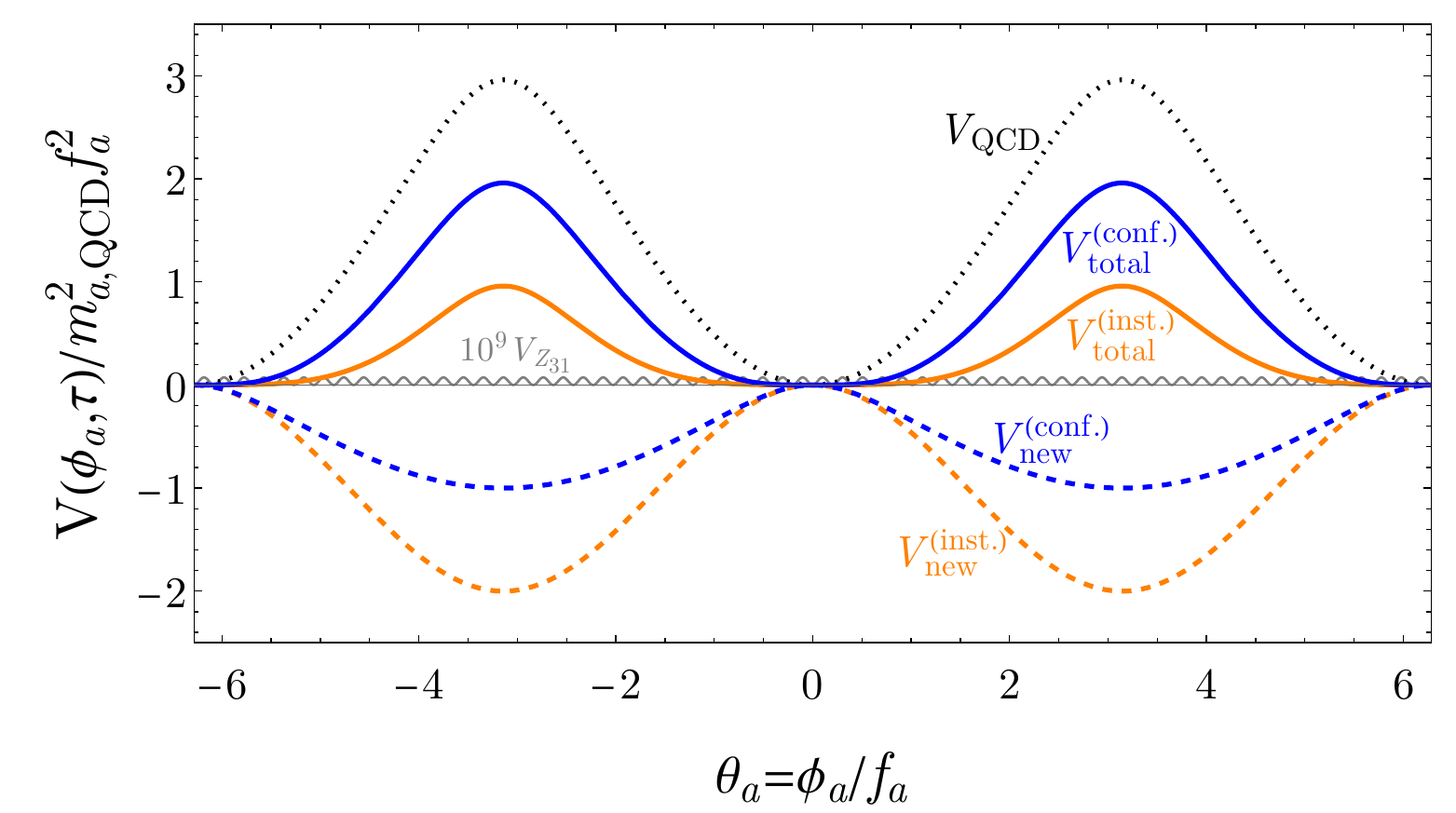}
    \caption{The total axion potential (solid) obtained from summing the QCD potential (dotted) with two examples of the potential from the mirror sector (dashed) which are the QCD-like potential (blue) and the instanton cosine potential (orange) with $\varepsilon \ll 1$.
    The solid grey line is the axion potential in the $\mathbb{Z}_N$ model~\cite{Hook:2018jle, DiLuzio:2021gos} with $N = 31$, where the magnitude has been amplified by a factor of $10^9$ for illustration purposes. Note that each potential has been shifted to be zero at~$\theta_a = 0$.}
    \label{fig:potential}
\end{figure}

We have given two examples of how a tuned axion potential with a light QCD axion can arise from a mirror sector where the tuning parameter depends on the mirror scalar field VEVs. 
The axion potentials for the two examples given in Section~\ref{sec:examples} are illustrated in Fig.~\ref{fig:potential}.
It is clear that a precise cancellation only happens at the origin and not at other $\theta_a$ values. For instance, there is still a large, untuned potential barrier near $\theta_a=\pi$, where the amplitude of the total potential is not suppressed by $\varepsilon$, and only near the origin does the total potential become much flatter. This is in contrast to the $\mathbb{Z}_N$ model studied in Ref.~\cite{DiLuzio:2021gos}. 

This qualitative form of the axion potential motivates studying the cosmological evolution and in particular, the effects on the relic dark matter abundance that have been previously unexplored. For simplicity, we will consider the cosine form of the potential Eq.~\eqref{eq:analVinst} to study the cosmological evolution in Section~\ref{sec:cosevol}, although we expect the qualitative features to remain the same for the QCD-like potential Eq.~\eqref{eq:potential_new_1}. The cosmological evolution will be studied for the range $10^{-8} \lesssim \varepsilon \le 1$, corresponding to the phenomenologically relevant region,  which represents a tuning in the potential down to $10^{-8}$. Furthermore, we will remain agnostic about an explicit mechanism that could explain this tuning\footnote{In the $\mathbb{Z}_N$ model~\cite{Hook:2018jle,DiLuzio:2021gos}, a small $\varepsilon$ does not require tuning, which in the large $N$ limit is approximately given by 
$\varepsilon \approx \brr{\dfrac{1-z^2}{\pi}}^{1/4} \brr{1+z}^{1/2} N^{3/4} z^{(N-1)/2}$,
where $z \equiv m_u/m_d$. However, our analysis does not immediately apply to this model because the potential amplitude also becomes suppressed (as seen in figure~\ref{fig:potential}) and affects the hilltop misalignment calculation.} since, as will be shown, the form of the potential in Figure~\ref{fig:potential} leads to an interesting cosmological evolution. 

\subsection{Finite temperature effects}
\label{sec:thermal}

Next we consider the finite temperature effects on the axion potential as well as constraints on the mirror sector in the early universe. First, we have assumed a complete mirror sector which has the same particle content as the SM. If the  new particles from the mirror sector are in thermal equilibrium with the SM sector they will contribute to the number of relativistic degrees of freedom in the early universe thereby changing the number of effective neutrino species $N_{\rm eff}$. The measured value of $N_{\rm eff}$ from Big Bang Nucleosynthesis (BBN) and the Cosmic Microwave Background (CMB) is very close to the SM prediction and consequently imposes very strong constraints on new physics parameters.

In order to satisfy the constraints on $N_{\rm eff}$, the mirror sector temperature must be lower than the SM plasma temperature. With a lower temperature, the mirror sector contributes less entropy density which helps to avoid the bound on $N_{\rm eff}$. It turns out that the temperature $T'$ of the mirror sector should satisfy~\cite{DiLuzio:2021gos}
\begin{equation}
\label{eq:BBN_CMB}
    \text{BBN: } \quad T' < 0.51 \, T_{\rm SM}\,,
    \quad\quad
    \text{CMB: } \quad T' < 0.60 \, T_{\rm SM} \,,
\end{equation}
where $T_{\rm SM}$ is the temperature of the SM thermal bath. We assume that $T'/T_{\rm SM}$ is constant throughout the whole radiation-dominated epoch and is much smaller than unity. 
Such a hierarchy in the plasma temperatures may be realized via the asymmetric reheating scenario~\cite{Berezinsky:1999az,Krauss:1985ac}. 

Furthermore, there can also be portal couplings between the mirror and SM sector~\cite{Berezhiani:2005ek,DiLuzio:2021gos}. In particular, the renormalizable Higgs coupling $\lambda_{H H'} H^\dagger H H^{\prime \dagger} H'$ must be sufficiently suppressed, in the first example discussed in Section~\ref{sec:Z2mirrorHiggs} where the mirror Higgs VEV $v'\lesssim v$. Other portal couplings, such as the kinetic mixing $\epsilon_{AA'} F_{\mu\nu} F^{\prime\mu\nu}$ between the hypercharge gauge bosons are also assumed to be sufficiently suppressed to avoid any cosmological constraints. For the $\mathbb{Z}_2$ model, the corresponding limits are $\lambda_{HH'},\epsilon_{AA'} \lesssim 10^{-8}$~\cite{Berezhiani:2005ek}.
Instead, in the example considered in Section~\ref{sec:spontbrokenmirrorQCD}, there is a large hierarchy between $v$ and $v'$. The portal couplings allow heavy particles in the mirror sector to decay to SM particles at high temperature, weakening the constraints.

Since we are interested in how the locally-flat axion potential modifies the misalignment mechanism in the early Universe, the temperature effects on the axion potential must also be taken into consideration. 
In what follows, we focus on the scenario where the mirror QCD potential is already present before the axion starts to oscillate. In the early universe this requires that at some time before the QCD phase transition the mirror sector temperature at least satisfies the condition $T' <v_S'$, so that the mirror sector potential dominates. In addition, we also require that $T'< T$ in the early universe so that the Hubble parameter is solely determined by $T$ and more specifically, after the QCD phase transition, the mirror sector temperature needs to satisfy Eq.~(\ref{eq:BBN_CMB}).
Initially the axion field is frozen and only when the Hubble constant becomes of order the axion mass scale, $m_{a,{\rm QCD'}}$, generated in the mirror sector, will the axion field start to oscillate around $\theta_a=\pi$. As the universe cools, the QCD phase transition eventually occurs at a later time, at which point the potential gradually becomes flat around the new minimum at the origin.

\begin{figure}
    \centering
    \includegraphics[width=0.8\linewidth]{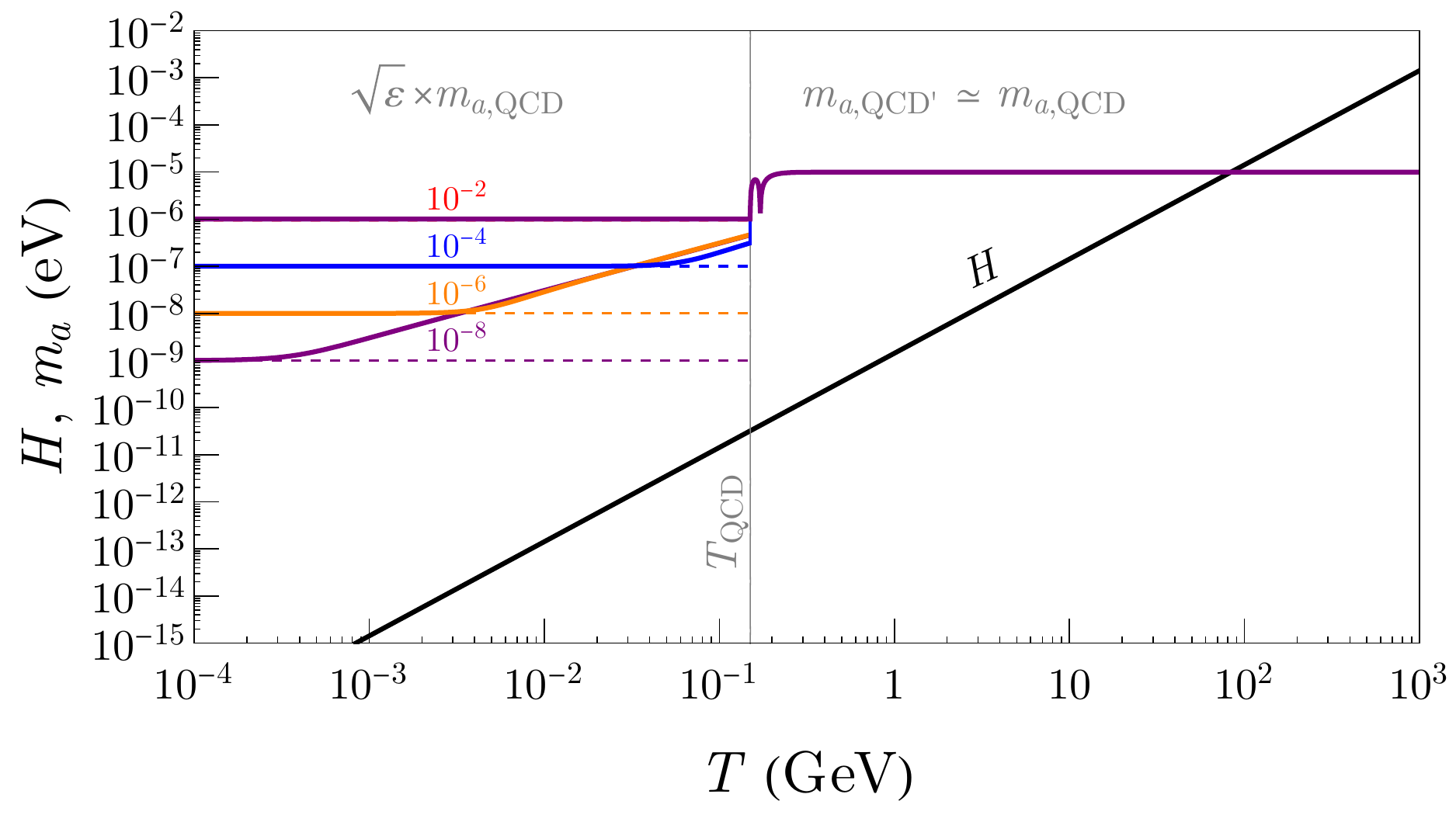}
    \caption{The evolution of the Hubble expansion rate $H$ and axion mass $m_a$ as a function of temperature $T$, assuming radiation domination in computing $H$. The axion mass, indicated by the colored lines (except for the diagonal part), is calculated by taking the second derivative of the potential Eq.~\eqref{eq:tem_dep_V} using Eq.~\eqref{eq:analVinst}, evaluated at the time-dependent minimum. The diagonal part represents the envelope of the oscillatory local mass (i.e.~second derivative evaluated at the axion oscillation amplitude). This corresponds to where the axion evolution is dominated by the quartic term, causing the difference between the local mass and the axion mass at the time-dependent minimum. The figure assumes $m_{a,\text{QCD}} = 10^{-5}$ eV and the values of $\varepsilon$ are indicated for the different colored curves.}
    \label{fig:H_ma}
\end{figure}

 As is well known for the QCD sector, at a temperature far above the QCD transition temperature $T_{\rm QCD}$, the dilute instanton approximation can well describe the axion dynamics. This gives rise to the temperature-dependent axion potential
 \begin{equation}
     V(\phi_a, T) = - m_a^2(T) f_a^2 \cos\brr{\dfrac{\phi_a}{f_a }}\,,
 \end{equation}
where the temperature dependent axion mass is given by~\cite{Gross:1980br} 
\begin{equation}
    m_a^2(T) \approx {\cal O}(10^{-2}) \times m_a^2 \brr{\dfrac{\text{GeV}}{T}}^8 \,,
\end{equation}
with the temperature scaling chosen to be consistent with the dilute instanton gas approximation. However, as the temperature becomes close to the QCD confinement scale, the temperature-dependent potential becomes analytically incalculable and only lattice simulations can reliably describe the behavior near the phase transition.
For temperatures, $T\ll T_{\rm QCD}$ the prediction Eq.~\eqref{eq:QCD_axion_potential} from chiral perturbation theory is a valid description. In the following, we follow~\cite{DiLuzio:2021gos} and assume that the temperature-dependent axion potential is given by
\begin{equation}\label{eq:dilute_instanton_V}
    V_{\rm QCD}(\phi_a,T) =  V_{\rm QCD}(\phi_a,\bar{\theta}) h(T) \,,
\end{equation}
where $V_{\rm QCD}(\phi_a,\bar{\theta})$ is given in Eq.~\eqref{eq:QCD_axion_potential} and the temperature factor $h(T)$ is defined to be
\begin{equation}
    h(T) = \left\{  
             \begin{array}{cc}  
              1 &  \qquad T \leq T_{\rm QCD}\,, \\  
              \brr{\dfrac{T_{\rm QCD}}{T}}^8 & \qquad T > T_{\rm QCD} \,. \\  
             \end{array}  
\right.  
\label{eq:hfactor}
\end{equation}
The interpolation choice in Eq.~\eqref{eq:hfactor} has several advantages. First, the potential is continuous across the phase transition temperature, $T_{\rm QCD}$. We ignore the finite temperature corrections when the plasma temperature is slightly lower than $T_{\rm QCD}$. Above $T_{\rm QCD}$, the potential should also be interpolated to match Eq.~\eqref{eq:dilute_instanton_V}, which will be ignored since it does not drastically affect the qualitative behavior. Therefore, the total axion potential can be parametrized as
\begin{equation}\label{eq:tem_dep_V}
    V_{\rm total}(\phi_a,T) = V_{\rm QCD}(\phi_a,T) + V_{\rm new}(\phi_a)\,.
\end{equation}
Note that we have ignored the temperature dependence of $V_{\rm new}$ since, as stated above, the new sector temperature $T'$ is assumed to already be lower than the mirror QCD confinement or spontaneous breaking scale when the axion dynamics starts. The evolution of the Hubble expansion rate and axion mass as a function of temperature is shown in Fig.~\ref{fig:H_ma}. When $H$ drops down to $\sim 3 m_a$, the axion field starts to roll. At the QCD phase transition, the Hubble constant is $H_{\rm QCD}\approx 10^{-11}$ eV. By construction, the axion mass contribution from the mirror sector is very close to that of QCD. For $m_a \gg H_{\rm QCD}$, the axion field starts to oscillate around $\theta_a = \pi$ before the QCD potential contributes. 
The axion energy density first scales as matter until the contribution from the SM QCD sector significantly modifies the axion potential. To calculate the axion dark matter abundance, one needs to solve the full equation of motion numerically.

\section{Cosmological Evolution}
\label{sec:cosevol}

\begin{figure}
     \centering
     \includegraphics[width=\linewidth]{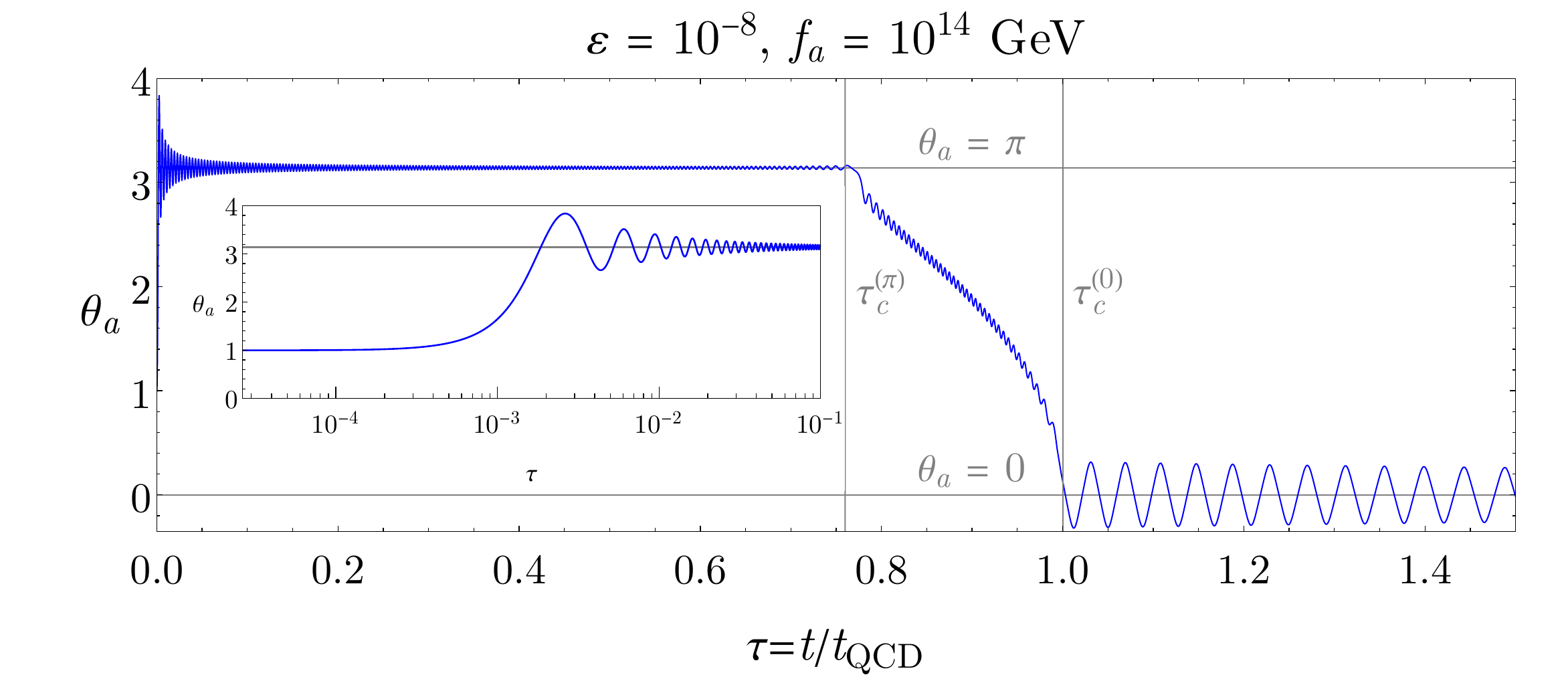}
     \caption{The evolution of the axion field $\phi_a = \theta_a f_a$ for $f_a = 10^{14}$ GeV and $\varepsilon = 10^{-8}$. The inset figure shows the evolution at early times with a logarithmic time scale when $\theta_a$ first relaxes to $\pi$ from an initial misalignment of $\theta_a=1$.}
     \label{fig:a_evolve_example}
 \end{figure}

Next we consider the cosmological evolution for the axion potential Eq.~\eqref{eq:tem_dep_V}.
Since the QCD potential in the SM sector is highly suppressed above $T_{\rm QCD}$, it is negligible until the temperature is sufficiently close to $T_{\rm QCD}$.  However, it is not reasonable to add the SM potential contribution abruptly, like a step function. 
As we will show in the following, the time scale matters and each step should be carefully tracked. Thus, the cosmological evolution before $T_{\rm QCD}$ is determined by the mirror axion potential which has its minimum at $\pi$, as shown in Fig.~\ref{fig:potential}, and then below $T_{\rm QCD}$ the minimum shifts to the origin, where there is a flattened potential.
Our primary task is to solve for the evolution of the axion field $\phi_a(t)$. Under the cosmic expansion dominated by the SM plasma, the equation of motion of the axion field is given by 
\begin{equation}
    \Ddot{\phi}_a + \dfrac{3}{2t} \dot{\phi}_a = -\dfrac{\partial V_{\rm total}(\phi_a,T)}{\partial \phi_a} \,,
    \label{eq:phiEOM}
\end{equation}
where the initial conditions are given in Section~\ref{sec:thermal} and we have used the explicit relation $H(t)=1/(2 t)$ for the radiation-dominated epoch. To obtain the axion relic abundance, we will begin the cosmological evolution just before $H\sim m_{a,\text{QCD}}$, and set the initial field value, $\theta_{a,0} \sim {\cal O}(1)$. For numerical simplicity, we convert Eq.~\eqref{eq:phiEOM} to an equation for $\theta_a (\equiv \phi_a/f_a)$ given by
\begin{equation}
    \theta_a'' + \dfrac{3}{2 \tau}  \theta_a' = -\dfrac{\partial V_{\rm total}(\theta_a,T)}{\partial \theta_a} \dfrac{1}{\brr{2 H_{\rm QCD}}^2 f_a^2}\,,
\end{equation}
where the prime $(')$ denotes the derivative with respect to the rescaled time $\tau = t/t_{\rm QCD}$. The evolution $\theta_a(t)$  for a particular choice of $(f_a,\varepsilon)$ is shown in Fig.~\ref{fig:a_evolve_example}. As can be seen in the figure, the evolution of the axion can be naturally divided into three stages. The first stage corresponds to $\tau < \tau_{c}^{(\pi)} \simeq 0.76$ (to be derived in Section~\ref{sec:A-NAevolution}), where the axion field starts to roll and oscillates around $\theta_a=\pi$ with a decreasing amplitude and a high frequency. After $\tau_{c}^{(\pi)}$, the minimum at $\pi$ begins to move towards the origin. The axion field tracks and oscillates around this shift of the minimum with a sizeable oscillation amplitude until $\tau$ is very close to one. 
For $\tau > 1$, the axion field keeps oscillating around the CP-conserving minimum at the origin with a large amplitude but much lower frequency that depends on the explicit parameter choice. 

To understand such a behavior, one needs to trace the time-dependent potential throughout the process. The following shows this analysis step by step.

\subsection{Adiabatic and non-adiabatic evolution}
\label{sec:A-NAevolution}

\begin{figure}
    \centering
   \includegraphics[width=0.8\linewidth]{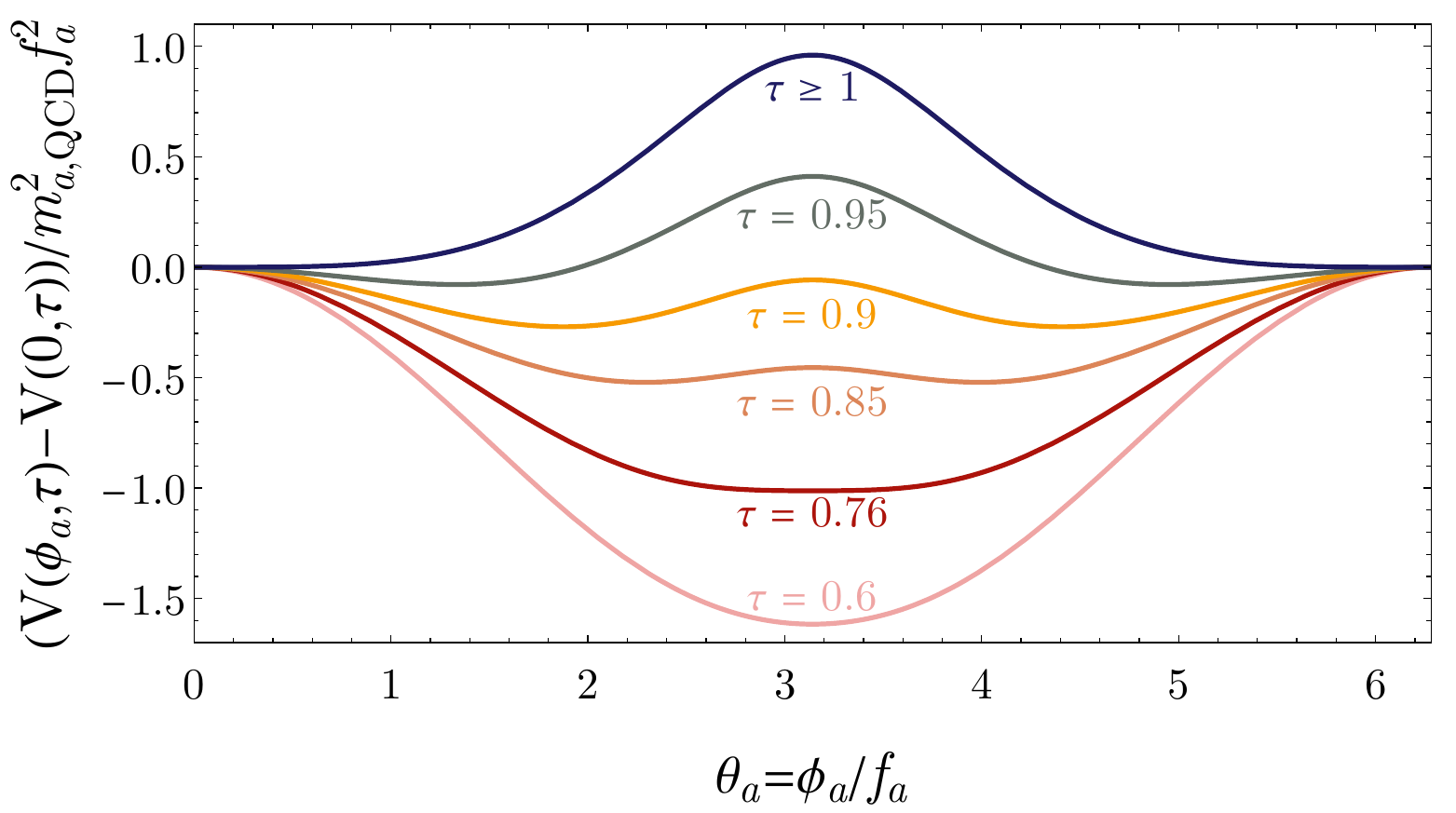}
    \caption{The evolution of the time-dependent potential $V_{\rm total}(\phi_a,\tau) - V_{\rm total}(0,\tau)$ in units of $m_{a,{\rm QCD}}^2 f_a^2$ at several $\tau$ values as the SM temperature approaches $T_{\rm QCD}$ (i.e., $\tau = t/t_{\rm QCD}\rightarrow 1$). Note that for each value of $\tau$ we have set the potential value at the origin to be zero.}
    \label{fig:time-evolv-V}
\end{figure}

As the universe cools and the SM plasma temperature approaches $T_{\rm QCD}$, the SM contribution to the total axion potential gradually increases. At $T_{\rm QCD}$, the minimum of the total potential occurs at the origin $\theta_a = 0$ and therefore during the evolution, the global minimum position must continuously change from $\theta_a = \pi$ to $\theta_a = 0$. To quantitatively describe such an evolution, we expand Eq.~\eqref{eq:tem_dep_V} around $\theta_a = \pi$ to quadratic order 
\begin{equation}
    V_{\rm total}(\phi_a,t) \approx \dfrac{1}{2} \brr{1- \varepsilon - 3 \tau^4} m_{a,{\rm QCD}}^2 f_a^2 \brr{\theta_a - \pi}^2 + \cdots 
    \label{eq:pipot}
\end{equation}
where we have converted the dependence on the temperature to the rescaled cosmic time $\tau$ using the relation $\tau = (T/T_{\rm QCD})^{-2}$ (we ignore the change of degrees of freedom for simplicity). 
For early times, when $t \ll t_{\rm QCD}$ (or $\tau\ll 1$), this is just the leading expansion of the cosine potential with the axion mass equal to the QCD axion mass times the $1-\varepsilon$ factor. As the temperature drops and $\tau$ increases, the $\tau^4$ term in Eq.~\eqref{eq:pipot} becomes larger, causing the potential near $\pi$ to become flatter. In the case of $\varepsilon \ll 1$, the potential Eq.~\eqref{eq:pipot} then flips sign when 
$\tau= \tau_{c}^{(\pi)} \approx 3^{-1/4} \simeq 0.76$. This behavior is depicted in Fig.~\ref{fig:time-evolv-V} where the time-dependent potential is shown for different values of $\tau$. 
As can be seen in Fig.~\ref{fig:time-evolv-V}, the potential at  $\tau = \tau_{c}^{(\pi)} = 0.76$  becomes rather flat around $\theta_a = \pi$. As $\tau$ continues to increase, the $\theta_a = \pi$ local minimum turns into a local maximum, with new local minima appearing on either side of $\theta_a=\pi$. Due to the periodicity of the potential, we only consider axion field values $\theta_a\leq\pi$, where the local minimum moves to $\theta_a=0$ as $\tau$ increases beyond $\tau_{c}^{(\pi)}$.

The form of the potential implies that the axion field initially oscillates around $\theta_a = \pi$. 
As $\tau\rightarrow \tau_{c}^{(\pi)} = 0.76$, the effective oscillation frequency decreases accordingly until the mass-squared term changes sign at $\tau_{c}^{(\pi)} = 0.76$. 
As the local minimum shifts away from $\pi$ towards the origin, the axion field also adjusts itself and follows the new local minimum.
For $m_{a,\text{QCD}} \gg H_{\rm QCD}$, the axion field evolves adiabatically, oscillating around the new local minimum with a  local oscillation frequency equal to the second derivative of the potential at the local minimum.
For $\tau> \tau_{c}^{(\pi)}$, the local oscillating amplitude becomes visible and is amplified compared to that prior to $\tau_{c}^{(\pi)}$. 
Such an enhancement is due to the onset of non-adiabatic behavior, which is determined by the competition between how fast the local minimum shifts away from $\pi$
and the oscillation frequency.
Again, the high power of the time dependence causes the local minimum to quickly move from $\pi$. 
Since the axion field speed is suppressed by the tiny amplitude and the decreasing oscillating frequency as $\tau \rightarrow 0.76$, the axion field fails to keep up with the location of the new local minimum. As a result, when the axion field starts to adjust and oscillate around the new temporary minimum, the local minimum has already moved away. This gives rise to a short period of non-adiabatic evolution immediately after $\tau_c^{(\pi)}$.
This behavior can be seen near $\tau_{c}^{(\pi)}=0.76$ in Fig.~\ref{fig:a_evolve_example}. Later, the axion field evolves adiabatically, by catching up and oscillating around the time-varying location of the local minimum.  
This explains the enveloped oscillation pattern for the second stage during $0.76 < \tau < 1$ in Fig.~\ref{fig:a_evolve_example}.

As $\tau\rightarrow 1$, we need to take a closer look at the behavior near the origin. Analogously, we can expand the total potential around $\theta_a = 0$ to obtain
\begin{equation}\label{eq:expan_0}
    \frac{V_{\rm total}(\theta_a,\tau)}{m_{a,\rm QCD}^2 f_a^2}  = \dfrac{1}{2} \brr{-1 + \varepsilon + \tau^4} \theta_a^2  +  \dfrac{1}{24} \brr{1-\varepsilon - \dfrac{\tau^4}{3}} \theta_a^4 + \cdots
\end{equation}
Assuming $\varepsilon \ll 1$, when $\tau$ is very close to one, the quartic coefficient has the approximately constant value $1/36$, while the quadratic term flips sign at $\tau_{c}^{(0)} = \brr{1-\varepsilon}^{1/4} \simeq 1- \varepsilon/4$. 
Hence, for $\tau<\tau_{c}^{(0)}$, the origin is a local maximum and there is a nonzero local minimum located at
\begin{equation}\label{eq:vev_evolving}
    \theta_{\rm min}(\tau) = 3 \brr{1-\varepsilon - \tau^4}^{1/2} = 3 \brr{\tau_{c}^{(0) \, 4} - \tau^4}^{1/2} \approx 6 \brr{\tau_{c}^{(0)} - \tau}^{1/2} \,.
\end{equation}
Taking the time derivative of Eq.~\eqref{eq:vev_evolving}, we obtain $\theta_{\rm min}' \sim \brr{\tau_{c}^{(0)} -\tau}^{-1/2}$. This means that as $\tau\rightarrow \tau_{c}^{(0)}$, the local potential minimum moves towards the origin at a faster and faster rate, which leads to non-adiabatic evolution. For $\tau>\tau_{c}^{(0)}$, when the origin becomes the local minimum, the axion field is still located at a nonzero value where it was previously oscillating around a nonzero local minimum. This nonzero value, therefore, provides the initial amplitude for the final oscillation phase around the origin. 
The initial misalignment angle $\theta_{a,0}$ can significantly affect the magnitude of this initial amplitude for the final oscillation stage around the origin. The final axion dark matter relic abundance crucially depends on this initial amplitude which will be computed in the next subsection.

\subsection{Axion relic abundance}

To calculate the axion dark matter relic abundance today, we need to obtain the value of the axion field amplitude immediately after $t_{\rm QCD}$, denoted as $\theta_{A,0}$. The axion field evolution has qualitatively different characteristics depending on the value of the decay constant $f_a$, and the tuning parameter $\varepsilon$. In the following three subsections, we will consider several regimes for the decay constant and discuss them separately. 

We first discuss the potential and evolution that is in common between all the different regimes. As the time approaches $\tau = 1$ and thereafter, the potential \eqref{eq:expan_0} is approximated as
\begin{equation}
   \dfrac{V_{\rm total}(\theta_a)}{m_{a,\rm{QCD}}^2 f_a^2} \approx \dfrac{1}{2} \varepsilon \theta_a^2 + \dfrac{1}{24} \brr{\dfrac{2}{3} - \varepsilon} \theta_a^4 \approx 
   \dfrac{1}{2} \varepsilon \theta_a^2 + \dfrac{1}{36}  \theta_a^4\,.
\end{equation}
The second derivative of this potential is 
\begin{equation}
\label{eq:ma}
   \dfrac{d^2 V_{\rm total}(\theta_a)}{d\phi_a^2} =  \dfrac{d^2 V_{\rm total}(\theta_a)}{f_a^2 d \theta_a^2} = m_{a,{\rm QCD}}^2 \brr{\varepsilon + \dfrac{1}{3} \theta_a^2 } \,,
\end{equation}
which implies that when the amplitude $\theta_A \gg \sqrt{\varepsilon}$, the axion field oscillates in a quartic potential, while for $\theta_A \ll \sqrt{\varepsilon}$, the axion field instead oscillates in a quadratic potential. In the quartic potential, the axion oscillation amplitude evolves as $\theta_A(t) \sim \tau^{-1/2}$, while for the quadratic potential, the axion amplitude evolves as $\theta_A(t) \sim \tau^{-3/4}$. 
Thus, as the axion field $\theta_A$ evolves from $\theta_A \gg \sqrt{\varepsilon}$ to $\theta_A \ll \sqrt{\varepsilon}$, the power of the time dependence gradually changes from $-1/2$ to $-3/4$.

\subsubsection{Large $f_a$}
\label{sec:large_fa}

For sufficiently large values of $f_a$,  
the QCD axion mass can be smaller than the Hubble value at the QCD phase transition. This means the axion field is frozen at $t_{\rm QCD}$.
The minimum value of the decay constant for which this occurs is determined from the condition $m_{a,{\rm QCD}} = 3 H_{\rm QCD} \simeq 4 \times 10^{-11}$ eV leading to the critical value $f_{a,c} \simeq 2 \times 10^{17}$ GeV where the axion mass $m_a\approx m_{a,{\rm QCD}}$ near $t_{\rm QCD}$. Thus, for large $f_a> f_{a,c}$, the axion field only starts to evolve when the Hubble value is approximately equal to the local mass~(\ref{eq:ma}) evaluated at $\theta_{a,0}$ i.e. $m_a(\theta_{a,0}) = \sqrt{\varepsilon + \frac{1}{3} \theta_{a,0}^2} \, m_{a,\text{QCD}}$.
In this case, the axion relic abundance today is approximately given by
\begin{equation}
    \Omega_a \sim \dfrac{f_a^2 \sqrt{\varepsilon m_{a,\text{QCD}}}}{M_{\rm pl}^{3/2} T_{\rm eq}} \dfrac{\frac{1}{2} \varepsilon \theta_{a,0}^2 + \frac{1}{36} \theta_{a,0}^4}{\left(\varepsilon + \frac{1}{3} \theta_{a,0}^2\right)^{5/4}}\,,
    \label{eq:Omegalargefa}
\end{equation}
where $M_{\rm pl}$ is the Planck scale and $T_{\rm eq} \sim$ eV is the temperature at matter-radiation equality. The expression \eqref{eq:Omegalargefa} is derived by first computing the axion number density at the onset of the oscillations using the initial mass $n_a (T_{\rm osc}) = V_{\rm total}(\theta_{a,0})/m_a(\theta_{a,0})$ and then obtaining the energy density at matter-radiation equality $\rho_a(T_{\rm eq}) = m_a n_a(T_{\rm eq})$  using the vacuum mass $m_a$, where $n_a(T_{\rm eq}) = n_a (T_{\rm osc}) (T_{\rm eq}/T_{\rm osc})^3$.
To explain the observed dark matter abundance of $\Omega h^2 \simeq 0.12$, one can see that either $\varepsilon$ or $\theta_{a,0}$ has to be much less than unity, while for $m_{a,{\rm QCD}} = 4 \times 10^{-11}$ eV, $\varepsilon \lesssim 0.1$ is excluded by superradiance effects~\cite{Baryakhtar:2020gao,Unal:2020jiy,Mehta:2020kwu} and the gravitational wave event GW170817~\cite{Zhang:2021mks}. Alternatively, assuming $\varepsilon \gtrsim 0.1$ to satisfy the astrophysical constraints, we find $\theta_{a,0} \lesssim \mathcal{O}(10^{-3})$.
Therefore, the large $f_a$ scenario nearly reduces to the conventional QCD axion case because the mass tuning $\varepsilon$ is constrained to be nearly absent and the tuning in the misalignment angle is severe. We will not study this regime in further detail.

\subsubsection{Intermediate $f_a$}

\begin{figure}[t]
    \centering
    \includegraphics[width=\linewidth]{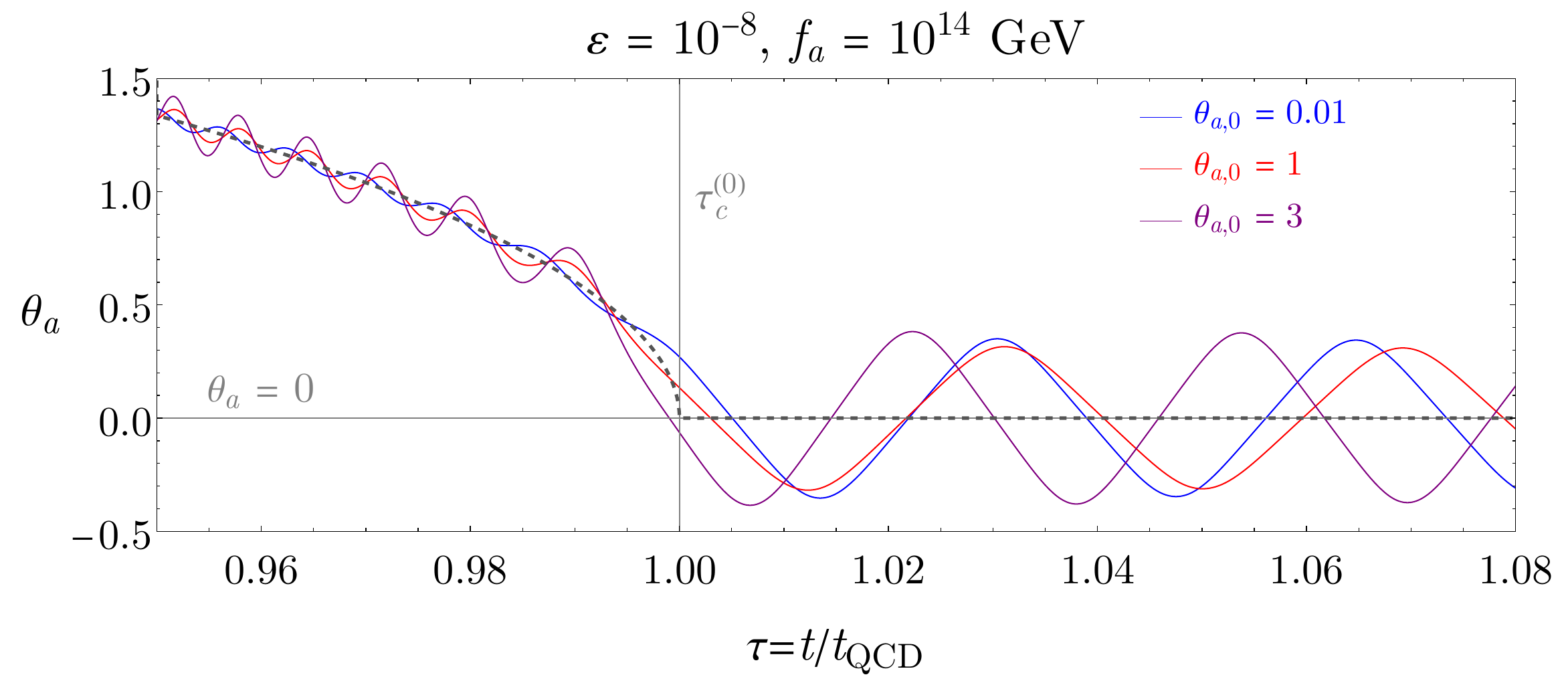}
    \caption{The evolution of the QCD axion field near $t_{\rm QCD} (\tau=1) $ for $\varepsilon = 10^{-8}$ and $f_a = 10^{14}$ GeV with three different initial angles $\theta_{a,0}$. The dashed line represents the evolution of the time-dependent local minimum $\theta_{\rm min}(\tau)$ and the gray vertical line denotes $\tau_{c}^{(0)}$ when the potential minimum reaches the CP-conserving origin.}
    \label{fig:intf_bench}
\end{figure}

For $f_a<f_{a,c}$, the axion field already starts to evolve before $t_{\rm QCD}$ and the equation of motion can be numerically solved until some time after $t_{\rm QCD}$. However, for sufficiently low $f_a$, even the tuned axion mass is rather high. This leads to a higher oscillation frequency, and the numerical computation starts to be time consuming. 
In this subsection, we restrict the axion decay constant to $10^{12}~{\rm GeV} < f_a < f_{a,c}$ and refer to these values as the intermediate $f_a$ regime.
The $\theta_{A,0}$ values in this range can become ${\cal O}(0.1)$, which means the axion evolution 
starts with a radiation-like scaling and a small $\varepsilon$ is needed to reproduce the correct dark matter abundance. For smaller $f_a$ values, $\theta_{A,0}$ would be more suppressed, leading to an initial matter-like scaling for higher $\varepsilon$ values, which will be discussed in Section~\ref{sec:lowfa}.
For example, the axion evolution is shown in Fig.~\ref{fig:intf_bench} for $\varepsilon = 10^{-8}$ and $f_a = 10^{14}$ GeV  with three initial values, $\theta_{a,0} = 0.01, 1$ and 3. For the chosen $\varepsilon$ value, $\tau_{c}^{(0)}$ is very close to one and cannot be distinguished from one in the figure.
The dashed line represents the evolution of the time-dependent local minimum $\theta_{\rm min}(\tau)$.
As seen in the figure, for $\tau$ not so close to $\tau_{c}^{(0)}$, the axion field keeps track of and oscillates around the time-varying minimum position $\theta_{\rm min}(\tau)$ as calculated in Eq.~\eqref{eq:vev_evolving}. 
Only as $\tau$ approaches very close to $\tau_{c}^{(0)}$, does the adiabatic condition break down.

Upon obtaining the $\theta_{A,0}$ value, identified as the axion field amplitude immediately after $t_{\rm QCD}$,
we need to further evolve the axion field to later times in order to compute the current relic abundance of the axion dark matter. This numerical evolution is time consuming and the computational uncertainty would accumulate to destabilize the convergence. However, there are alternative ways to obtain the final abundance at later times which we next discuss.

In principle, one can utilize the same method as in Section~\ref{sec:large_fa}, where the number density is computed using $\theta_{A,0}$ and dilutes with the cosmic expansion if the evolution is adiabatic. However, the small upper bound on $\theta_{a,0}$ derived in Section~\ref{sec:large_fa} can be relaxed with intermediate $f_a$ by allowing $\varepsilon \ll 1$ to obtain the observed dark matter abundance. 
This immediately implies that the period of evolution governed by the quartic term will be prolonged. The evolution is no longer adiabatic due to parametric resonance as we will elaborate in Section~\ref{sec:PR}. We leave the effects of parametric resonance for future work. Here we will continue to focus on estimating the abundance in the linear analysis.

\begin{figure}[t!]
    \centering
   \includegraphics[width=0.8\linewidth]{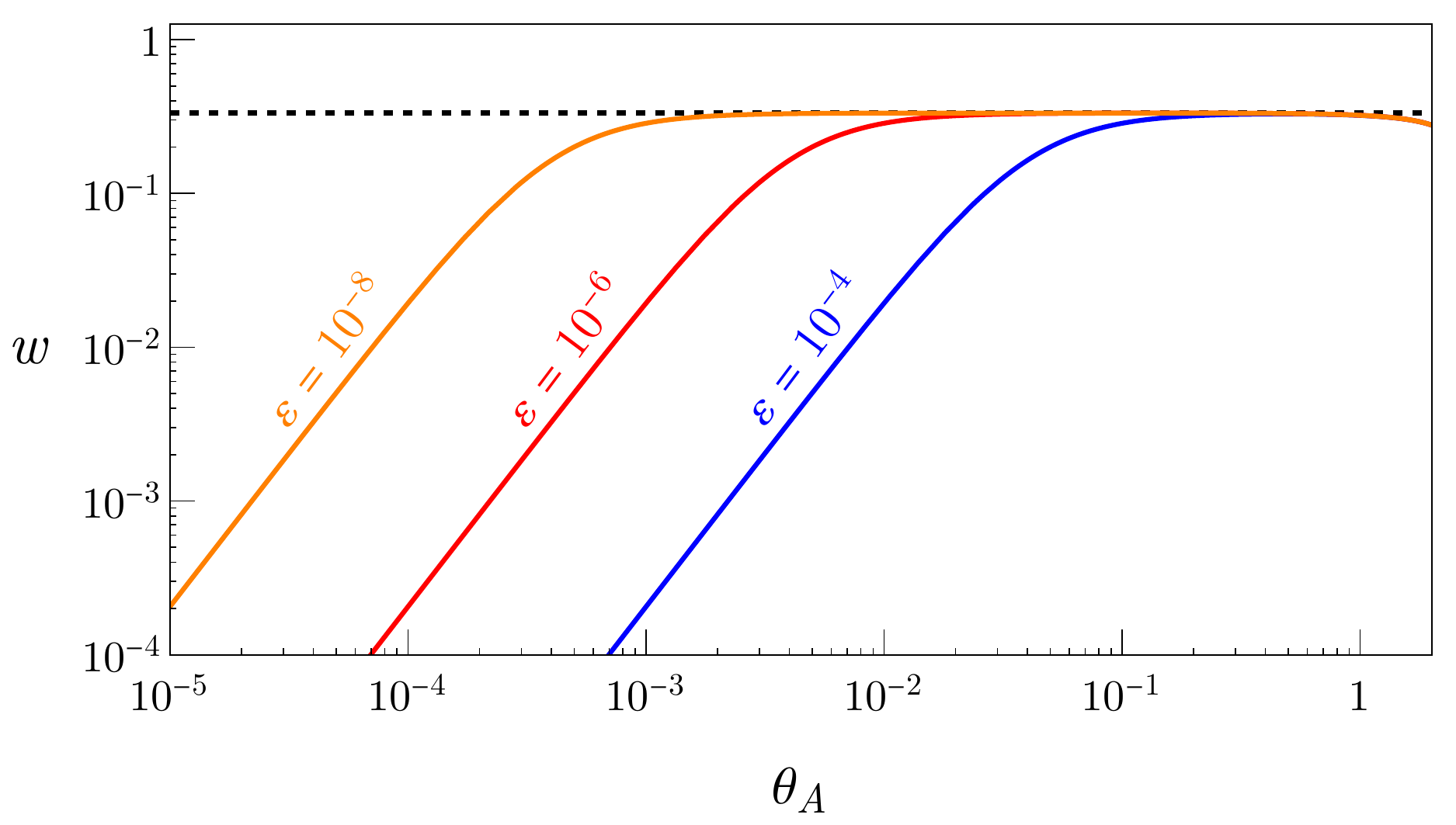}
    \caption{The equation of state parameter $w$ as a function of the axion oscillation amplitude~$\theta_A$, for various values of $\varepsilon$. As $\theta_A$ decreases from unity, the equation of state parameter transitions from radiation $w=1/3$ (dashed black line)  to matter $w=0$.}
    \label{fig:w-theta}
\end{figure}

In order to accurately capture the smooth transition from radiation-like to matter-like scaling of the axion energy density, we can consider the evolution of the axion field energy density, $\rho_a(t)$. From the Lagrangian, we know that
\begin{equation}
    \rho_a(t) = \dfrac{1}{2} f_a^2 \dot{\theta}_a^2 + V(\theta_a)\,.
\end{equation}
On the other hand, the axion field energy density decreases due to the cosmic expansion and therefore satisfies the scaling behavior
\begin{equation}\label{eq:rho_scale}
    \dfrac{\rho(t+\Delta t)}{\rho(t)} = \brr{\dfrac{a(t+\Delta t)}{a(t)}}^{-3 \gamma}\,,
\end{equation}
where $a(t)$ is the cosmic scale factor and the exponent $\gamma\equiv 1+w$, for the equation of state parameter $w$, is given by~\cite{Turner:1983he} 
\begin{equation}
    \gamma(\theta_A) = 2\dfrac{\int_0^{\theta_A} d\theta_a \srr{1-V(\theta_a)/V(\theta_A)}^{1/2} }{\int_0^{\theta_A} d\theta_a \srr{1-V(\theta_a)/V(\theta_A)}^{-1/2} }  \,.
    \label{eq:gammadef}
\end{equation}
The integration in Eq.~\eqref{eq:gammadef} is over one oscillation cycle 
where $\theta_A$ denotes the oscillation amplitude.
The Hubble friction damps the oscillation  and thus the oscillation amplitude gradually decreases. 
Expanding Eq.~\eqref{eq:rho_scale}, we obtain
\begin{equation}\label{eq:eom_theta_A}
    \dfrac{\dot{\theta}_A}{\rho} \dfrac{\partial \rho}{\partial \theta_A} = - 3 H \gamma(\theta_A) \,,
\end{equation}
where the dot ($\cdot$) refers to a time derivative. By solving Eq.~\eqref{eq:eom_theta_A}, we obtain the evolution of the oscillation amplitude $\theta_A(t)$, which can then be substituted into $V(\theta_A)$ to obtain the energy density $\rho_a(t)$ since the kinetic term is zero at $\theta_A$. This method greatly reduces the computation time. 
In Fig.~\ref{fig:w-theta}, we show the equation of state parameter, $w$ as a function of the axion field amplitude $\theta_A$ for various values of $\varepsilon$.
When $\theta_A \lesssim \sqrt{\varepsilon}$, the value of $w$ has significantly decreased from $w=1/3$ and as $\theta_A\rightarrow 0$, the value of $w$ also goes to zero corresponding to matter-like behavior.

After solving for the evolution $\theta_A(t)$, we can obtain
the axion dark matter relic abundance today from the equation
\begin{equation}\label{eq:Omega_a}
    \dfrac{\Omega_a}{\Omega_{m}} = 
    \dfrac{V(\theta_A(t_m))}{3 M_{\rm pl}^2 H(t_m)^2}   \brr{ \dfrac{H(t_m)}{H_{\rm eq}} }^{1/2}\,,
\end{equation}
where $\Omega_m$ is the current dark matter density and $t_m$ corresponds to the time when the axion energy density scales like matter.

\subsubsection{Low $f_a$}
\label{sec:lowfa}

When $f_a \le 10^{12}$ GeV, corresponding to the low $f_a$ case,
the amplitude $\theta_{A,0}$ is more suppressed compared to the intermediate $f_a$ case. This is due to the following two effects.

First, with the corresponding higher mass or oscillation frequency associated with low $f_a$, the axion field can more efficiently track the change in the location of the minimum. 
In particular, just before $\tau_{c}^{(0)}$, the higher frequency means that the axion can track the changing local minimum until it is much closer to $\tau_{c}^{(0)}$ compared to case with intermediate values of $f_a$. In other words, the axion angle $\theta_a$ now has a smaller amplitude when it begins to oscillate around the origin after $\tau_{c}^{(0)}$.

Secondly, for sufficiently small values of $f_a$, there is now a large hierarchy between the QCD axion mass $m_{a,{\rm QCD}}$ and $H_{\rm QCD}$. As derived in Section~\ref{sec:A-NAevolution}, the quadratic term in Eq.~\eqref{eq:expan_0} flips sign at $\tau_{c}^{(0)} \simeq 1- \varepsilon/4$. The time scale with respect to $t_{\rm QCD}$ can be characterized by $H_{\rm QCD}/\varepsilon$, while the axion mass, $m_a = \sqrt{\varepsilon} \, m_{a,\text{QCD}}$. Consequently, it is possible that $m_a \gg H_{\rm QCD}/\varepsilon$ for some $\varepsilon$ values, causing the axion to undergo high frequency oscillations around the origin during the time interval $\tau_{c}^{(0)} < \tau <1$. Starting from the time characterized by $\tau_{c}^{(0)}$, the coefficient of the quadratic term in the potential continuously evolves from $0$ to $\varepsilon$, while the coefficient of the quartic term is approximately constant at $1/36$. Moreover, the axion amplitude $\theta_A(\tau_c^{(0)})$ can already be smaller than $\sqrt{\varepsilon}$ so that it is the quadratic term that dominates the potential.
The total potential is therefore becoming steeper as the quadratic coefficient increases, which causes the axion field value to be further reduced between $\tau_c^{(0)}$ and $\tau=1$. 

\begin{figure}[t]
    \centering    \includegraphics[width=\linewidth]{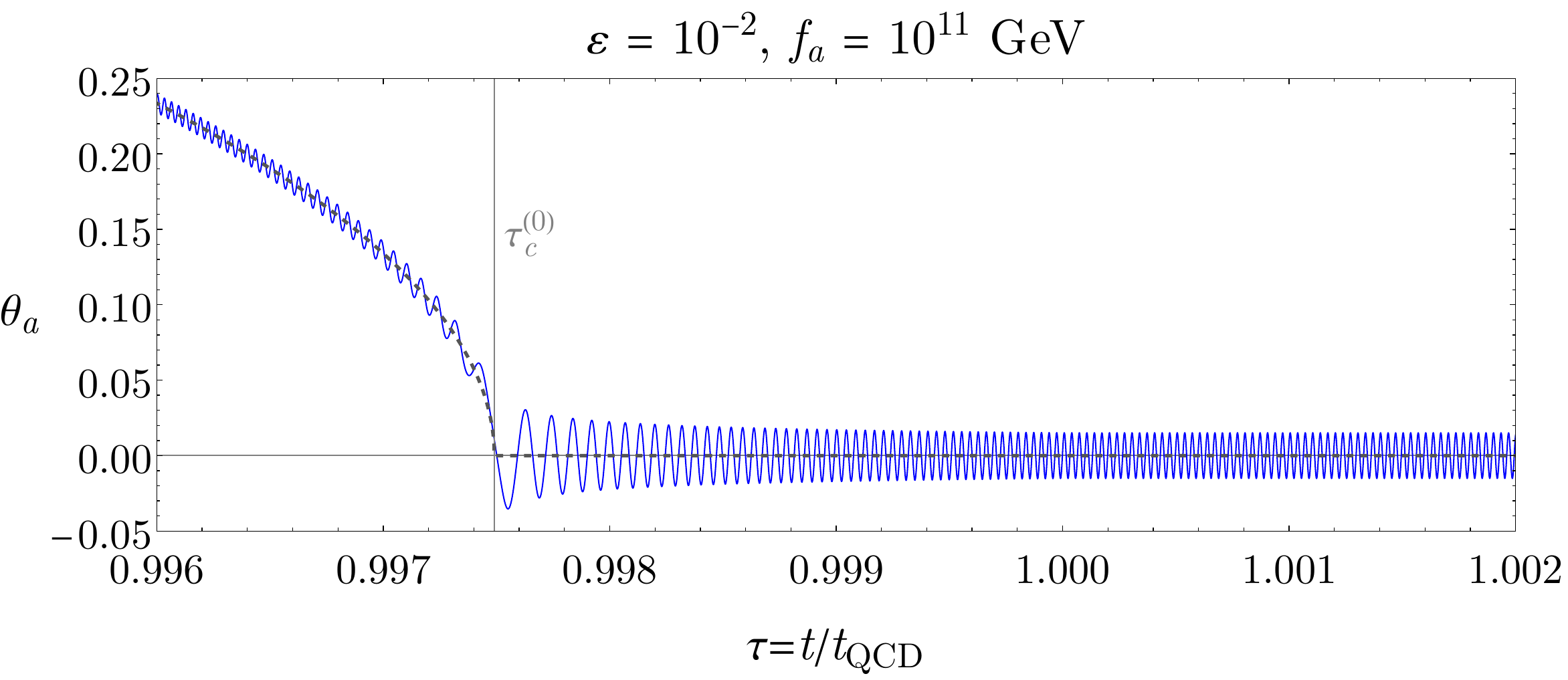}
    \caption{The axion field evolution for $\varepsilon = 10^{-2}$ and $f_a = 10^{11}$ GeV. The gray vertical line denotes $\tau_{c}^{(0)}$ when the potential minimum is at the CP-conserving origin. The dashed black line represents the position of the time-varying local minimum.}
    \label{fig:low_fa_1}
\end{figure}

We show an example of the axion evolution in Fig.~\ref{fig:low_fa_1} for the values $\varepsilon = 0.01$ and $f_a = 10^{11}$~GeV. 
As seen in the figure, the axion field adiabatically tracks the location of the minimum (indicated by the dashed line) until it is very close to $\tau_{c}^{(0)}$. Therefore, this causes the initial amplitude at $\tau_{c}^{(0)}$ for oscillations around the origin to be much reduced compared to cases with larger $f_a$. 
Furthermore, Fig.~\ref{fig:low_fa_1} also depicts the evolution from $\tau_{c}^{(0)}$ to $\tau = 1$. Since the coefficient of the quadratic term in the potential is increasing from  $\tau_{c}^{(0)}$, the axion field value at $\tau = 1$ decreases to approximately a factor of two smaller than its value right after $\tau_{c}^{(0)}$. 
Such an extra suppression is absent at higher values of $f_a$ or lower values of $\varepsilon$. 

Thus, taking into account the above two effects, 
the initial axion field amplitude $\theta_{A,0}$ at $t_{\rm QCD}$ is much smaller compared to the intermediate $f_a$ case. The axion relic abundance is still given by Eq.~\eqref{eq:Omega_a}, but now, with a lower value of $\theta_{A,0}$, the correct dark matter abundance can be obtained for a higher value of $\varepsilon$.

\subsubsection{Parametric resonance}
\label{sec:PR}

In the previous subsections, we focused on the coherent oscillations of the zero mode of the axion field. However, the fluctuation modes can be greatly amplified if the axion field oscillates in the non-quadratic potential. The classical axion field $\phi_a(t,\Vec{x})$ can be decomposed~as 
\begin{equation}
    \phi_a(t,\vec{x}) = \phi_a(t) + \delta \phi_a(t,\vec{x}) = \phi_a(t) + \brr{ \int \dfrac{d^3 k}{(2\pi)^3} b_k u_k(t) e^{i k x} + h.c.  }\,,
\end{equation}
where $b_k$ and $b_k^\dagger$ are the annihilation and creation operators, respectively, satisfying
\begin{equation}
    [b_k, b_{k'}^\dagger] = \brr{2\pi}^3 \delta^{(3)}\brr{k -k'} \ .
\end{equation}
Expanding the axion field equation of motion, the modes $u_k(t)$ satisfy
\begin{equation}
\label{eq:uk_evo}
    \Ddot{u}_k + 3 H \dot{u}_k + \srr{ \dfrac{k^2}{a^2} + V''(\phi_a)} u_k = 0\,,
\end{equation}
where $a$ is the cosmic scale factor. In the case of a high-power potential, $V''(\phi_a)$ is dependent on the explicit value of $\phi_a$, which could cause $V''(\phi_a(t))$ to highly oscillate with time.  This implies that the oscillation frequency of $V''(\phi_a(t))$ (the driving force) can scan over the modes and coincide with a particular momentum mode $k$. When this occurs, the solution of Eq.~(\ref{eq:uk_evo}) leads to an exponential growth or parametric resonance. 
Since the occupation number for such selected momenta is exponentially increased, this causes a large portion of the axion zero-mode energy to be converted to the quanta with specific momenta. 
This parametric resonance production would in turn deplete the oscillation of the homogeneous modes ($k=0$). Accordingly, the zero mode oscillation amplitude would be quickly reduced, which terminates parametric resonance. 

Generically, parametric resonance production occurs until the axion evolution is in the quadratic regime, namely $\theta_a \lesssim \sqrt{\varepsilon}$. The duration in the quartic regime may or may not be sufficiently long for fluctuations to be efficiently created and the zero mode to be depleted. Therefore, a numerical evaluation is warranted to determine the region of parameter space where parametric resonance is fully effective.
We numerically solved Eq.~\eqref{eq:uk_evo} and found that, for the $\varepsilon$ values required by the correct dark matter abundance, efficient parametric resonance occurs for $f_a \gtrsim 1.5\times 10^{11}$ GeV. The resonant momentum is slightly smaller than the initial mass of the axion field or $V''(\theta_{a,0})^{1/2}$. This means that the excited quanta are semi-relativistic and therefore one expects the parametric resonance production would change the final axion relic abundance by an ${\cal O}(1)$ factor. We do not perform a detailed calculation of the associated backreaction effects in parametric resonance. We expect that the allowed $(1/f_a,m_a)$ parameter space in our final result will only shift slightly towards the right in the figure due to a slight reduction of the number density. 

\begin{figure}
    \centering
    \includegraphics[width=0.8\linewidth]{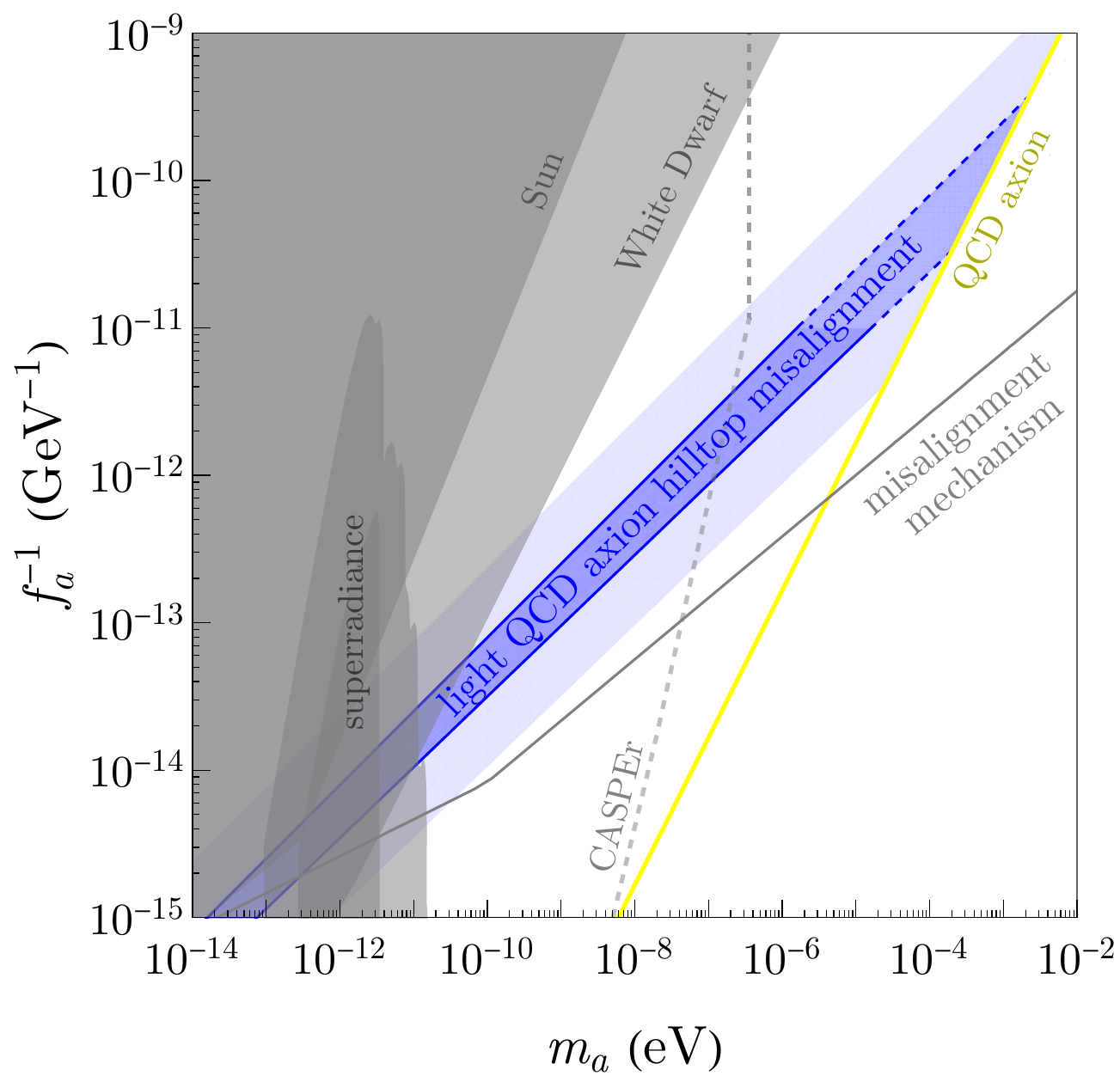}
    \caption{The predicted values of the axion mass $m_a$ and decay constant $f_a$ due to the hilltop misalignment mechanism. The darker blue band, bounded by the blue solid lines, shows the parameter space where the hilltop misalignment mechanism has been numerically confirmed to reproduce the observed dark matter abundance. The width of the band originates from varying the initial misalignment angle between $0.1 \pi \le \theta_{a,0} \le 0.9 \pi$. The blue band bounded by the dashed lines is obtained from extrapolating the solid lines without direct numerical confirmation due to the limited computation power. The lighter blue regions above and below the central blue band are obtained by shifting $f_a$ by factors of 1/3 and 3 from the central band, which estimates the expected uncertainty in the prediction due to parametric resonance. The gray shaded regions are excluded by observations as labeled, while the dashed gray line shows the projected sensitivity of CASPEr~\cite{Budker:2013hfa,JacksonKimball:2017elr}.}
    \label{fig:ma_fa}
\end{figure}

\subsubsection{Final results}
We summarize the final allowed parameter space in the $m_a$ versus $1/f_a$ plane in Fig.~\ref{fig:ma_fa}.
The yellow line refers to the conventional QCD axion mass prediction. 
The central blue band bounded by the solid blue lines depicts the values that give the correct dark matter relic abundance for initial misalignment  angles ranging from 0.1$\pi$ to $0.9\pi$. The central blue region is approximately given by
\begin{equation}
    8 \times 10^{-13} \GeV^{-1} \left( \frac{m_a}{10^{-8} \eV} \right)^{0.5} > f_a^{-1} > 3 \times 10^{-13} \GeV^{-1} \left( \frac{m_a}{10^{-8} \eV} \right)^{0.48} .
\end{equation}
The light blue region is extended from the blue band by a factor of $1/3$ and $3$ in $f_a$, to account for the uncertainty due to parametric resonance. 
Given the limit on computation time, we only scan for $f_a\lesssim 10^{11}$ GeV and simply extrapolate to lower values of $f_a$. The gray region is excluded by superradiance effects~\cite{Baryakhtar:2020gao,Unal:2020jiy,Mehta:2020kwu} and observations from the sun~\cite{Hook:2017psm,DiLuzio:2021pxd} and white dwarfs~\cite{Balkin:2022qer}. The dashed gray line refers to future projections of the CASPEr experiments. The solid gray line shows the usual misalignment prediction for an axion whose potential follows the same temperature dependence as the QCD axion in Eq.~(\ref{eq:hfactor}) and always has a minimum centered at $\theta_a = 0$ in contrast to our scenario where the minimum transitions from $\pi$ to $0$. In addition, along this solid gray line, we assume the axion mass is a parameter independent of $f_a$.
As seen in the figure, for $f_a\gtrsim 10^{14}$ GeV tuning parameter values satisfying $ \varepsilon \lesssim 10^{-8}$ have been excluded by the superradiance and white dwarf~\cite{Balkin:2022qer} observations. 
Thus, hilltop misalignment, favors $f_a\lesssim 10^{14}$ GeV corresponding to larger values of the tuning parameter $\varepsilon \gtrsim 10^{-8}$.
In particular, for $f_a \lesssim 10^{11}$~GeV, the axion mass is only one order of magnitude smaller than the QCD axion mass and therefore requires less tuning. 

\begin{figure}
    \centering
    \includegraphics[width=0.8\linewidth]{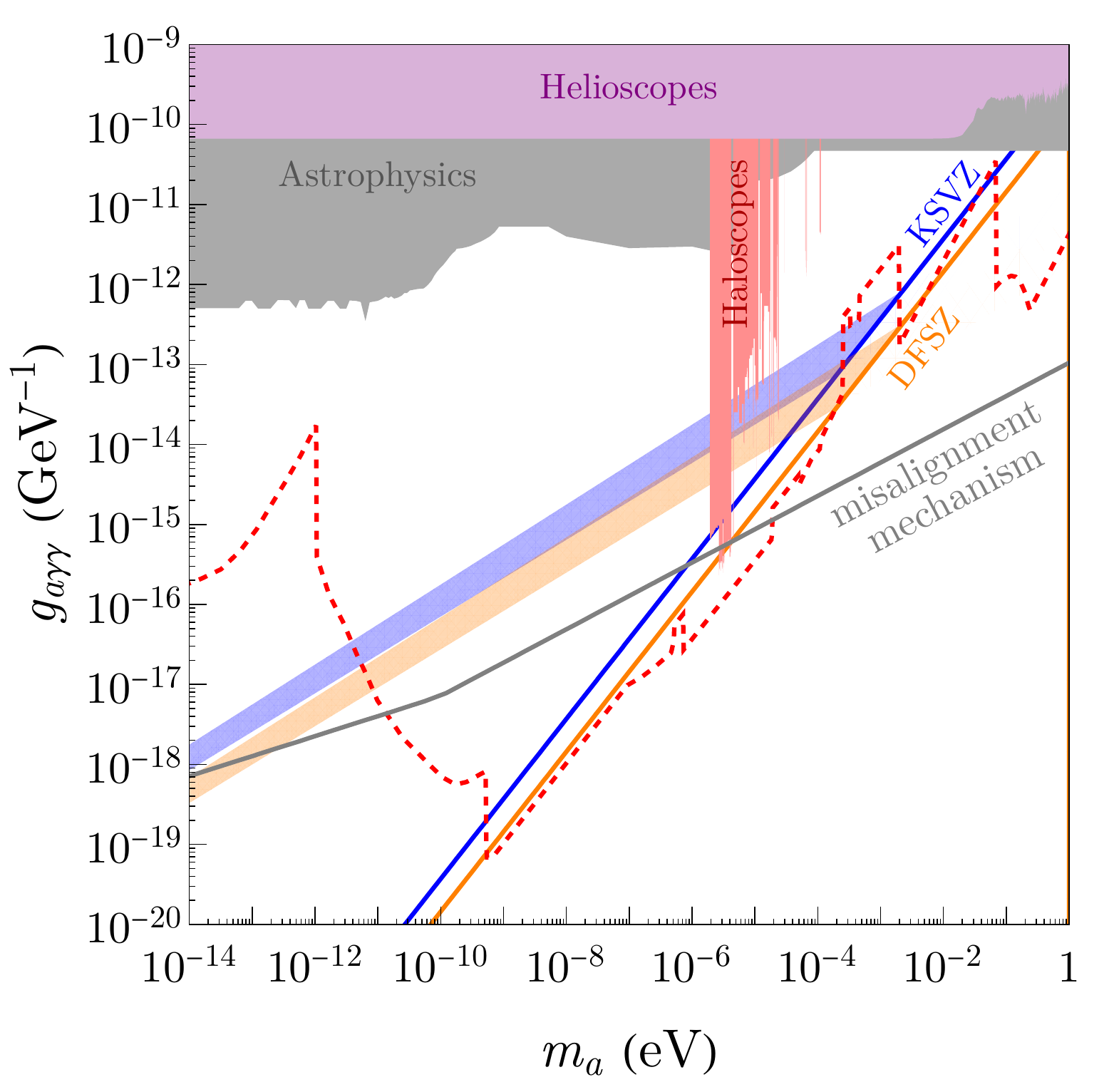}
    \caption{
    The predicted values of the axion mass $m_a$ and the axion-photon coupling $g_{a\gamma\gamma}$ due to the hilltop misalignment mechanism.
    Analogous to the blue band in Fig.~\ref{fig:ma_fa}, the blue (orange) band shows the parameter space where the hilltop misalignment mechanism numerically reproduces the observed dark matter abundance for the KSVZ (DFSZ) QCD axion model. The gray line shows the prediction of the misalignment mechanism from a QCD axion whose minimum is always located at $\theta_a=0$ as opposed to our scenario, where the minimum shifts from $\pi$ to $0$. The shaded regions at the top are excluded from different experimental searches as specified, while the dashed red line shows the projected sensitivity of future axion haloscope experiments.}
    \label{fig:ma_photon}
\end{figure}

In Fig.~\ref{fig:ma_photon}, we convert the constraint in the $(1/f_a, m_a)$ plane to the parametric $(g_{a\gamma\gamma}, m_a)$ plane (where $g_{a\gamma\gamma}$ is the axion-photon coupling) for the KSVZ and DFSZ axion scenarios, denoted in blue and orange, respectively. The colored band region depicts the allowed region from our analysis with the uncertainty due to parametric resonance and numerical precision. The haloscope experiments~\cite{Salemi:2021gck,Pandey:2024dcd,ADMX:2021nhd,ADMX:2021mio,Crisosto:2019fcj,Yang:2023yry,Kim:2023vpo,CAPP:2024dtx,Adair:2022rtw,Oshima:2023csb,Heinze:2023nfb,QUAX:2023gop,Ahyoune:2024klt,Gramolin:2020ict,DMRadio:2022pkf,Beurthey:2020yuq} have already excluded some of the parameter space for axion masses of ${\cal O}(10^{-6})$ eV as shown in red. The dashed line refers to the projected sensitivity from future haloscope experiments~\cite{DMRadio:2022pkf,Stern:2016bbw,Lawson:2019brd,Aja:2022csb,BREAD:2021tpx,Baryakhtar:2018doz}, which is expected to cover most of the allowed parameter space corresponding to $m_a > {\cal O}(10^{-12})$ eV. The regions excluded by helioscopes~\cite{CAST:2017uph} are indicated by purple and those by astrophysical searches~\cite{Ayala:2014pea,Dolan:2022kul,Reynes:2021bpe,Reynolds:2019uqt,Dessert:2022yqq,Noordhuis:2022ljw,Ning:2024eky} are indicated by gray. These curves are taken from Ref.~\cite{AxionLimits}.

\section{Conclusion}\label{sec:conclusion}

In this paper, we have studied the cosmological evolution of an axion lighter than the usual QCD axion where the strong CP problem is still solved. The axion potential is assumed to arise from a mirror sector whose $\bar\theta$ angle is shifted by $\pi$ compared to the SM sector. This allows the mirror contribution to the axion potential to be tuned against the usual QCD contribution, generating a lighter QCD axion. This gives rise to an axion potential that is much flatter near the CP-conserving minimum at the origin, while the axion potential at $\pi$ remains mostly unsuppressed.

Under the assumption that the temperature of the mirror sector remains lower than the  SM sector and any portal couplings between the two sectors are sufficiently suppressed, this distinctive form of the potential gives rise to a new, interesting axion evolution in the early universe. Much before the QCD phase transition, the axion potential is dominated by the mirror contribution which has a temporary, local minimum at $\pi$.   Assuming an order one initial axion value,  the frozen axion eventually rolls towards $\pi$ where it undergoes oscillations around that minimum. 
As the universe cools through the QCD phase transition, the QCD contribution to the axion potential becomes comparable to the mirror contribution causing the minimum at $\pi$ to shift to the origin. As a result of this shift, the axion gradually rolls from the newly-formed ``hilltop" at $\pi$, tracking and oscillating around the changing
local minimum as it heads towards the origin. However, just after $\pi$ and just before reaching the origin, there are brief periods of non-adiabatic behavior because the axion cannot keep up with the time-varying local minimum.
After the position of the local minimum reaches the origin, the lagging axion field position becomes the initial condition for oscillations around the origin.
This adiabatic behavior, accompanied by the brief non-adiabatic behavior, modifies the usual predictions from the conventional misalignment mechanism. In this ``hilltop" misalignment, the relic dark matter abundance can be obtained for $10^9 \lesssim f_a \lesssim 10^{14}$~GeV corresponding to the axion mass range $10^{-11} \lesssim m_a \lesssim 10^{-3}$~eV and a tuning of $10^{-8} \lesssim \varepsilon \lesssim 1$ in the axion mass. This differs from the usual parameter range for the conventional QCD axion and misalignment mechanism.

Our results rely on the distinctive form of the tuned axion potential. We have provided two examples of how such a potential could arise by relating the tuning to scalar field VEVs in the mirror sector that explicitly breaks the $\mathbb{Z}_2$ symmetry.  Nevertheless, it is worth further exploring how this tuning could be explained especially in relation to the Higgs mass hierarchy problem which has been ignored. For instance, in the $\mathbb{Z}_N$ model~\cite{Hook:2018jle, DiLuzio:2021gos} this tuning is explained by a large number of QCD copies although the resulting axion potential has many local minima and a suppressed amplitude at $\pi$. As the potential minimum transitions from $\theta_a = \pi$ to $N$ degenerate minima, the axion needs to go through different minima and ultimately settle down by chance to the right minimum at $\theta_a = 0$. In all of these models, the non-linear effects from parametric resonance as the axion field oscillates from the hilltop have been neglected. This would be interesting to further analyze in terms of the impacts on the dark matter abundance and gravitational waves. Specifically, observable gravitational waves might result from parametric resonance as shown in Refs.~\cite{Chatrchyan:2020pzh,Cui:2023fbg}.

In summary, despite the shortcomings of the underlying UV models, the axion potential in our ``hilltop" misalignment leads to an interesting cosmological evolution. This opens up the parameter region where the axion solves the strong CP problem and provides the missing dark matter component of the universe. This gives further motivation for experimental axion searches to explore a parameter region outside the conventional QCD axion predictions.

\section*{Acknowledgements}
We thank Keisuke Harigaya for useful discussions. This work is supported in part by the Department of Energy under Grant No.~DE-SC0011842 at the University of Minnesota and under Grant No.~DE-SC0025611 at Indiana University. For facilitating portions of this research, Z.L and K.F.L wish to acknowledge the Center for Theoretical Underground Physics and Related Areas (CETUP*), The Institute for Underground Science at Sanford Underground Research Facility (SURF), and the South Dakota Science and Technology Authority for hospitality and financial support, as well as for providing a stimulating environment when this work was finalized. T.G. and Z.L. also acknowledge the Aspen Center for Physics, where parts of this work were performed, which is supported by National Science Foundation grant PHY-2210452.

\bibliography{references}

\end{document}